\documentclass[12pt,reprint,noshowpacs,nofootinbib,notitlepage,amsmath]{revtex4-1}
\usepackage{setspace}
\allowdisplaybreaks
\usepackage{graphicx,color}
\usepackage[colorlinks=true,citecolor=blue,linkcolor=blue,urlcolor=blue]{hyperref}
\usepackage[charter]{mathdesign}
\usepackage{multirow}
\usepackage{enumerate}
\usepackage{slashed}

\DeclareSymbolFontAlphabet{\mathcal}{symbols}
\DeclareSymbolFont{symbols}{OMS}{xmdcmsy}{m}{n}
\DeclareSymbolFont{largesymbols}{OMX}{xmdcmex}{m}{n}
\SetSymbolFont{symbols}{bold}{OMS}{xmdcmsy}{b}{n}
\begin{document}
\title{\color{blue}\Large A ghost and a naked singularity; facing our demons}
\author{Bob Holdom}
\email{bob.holdom@utoronto.ca\\This paper is a write-up of a talk I gave at the CERN workshop ``Scale Invariance in Particle Physics and Cosmology'', Jan 28 - Feb 1, 2019. The talk builds on work in collaboration with Jing Ren, but I give some different perspective, as in Section II, and some new results (the new solution in Section IV and analysis of more LIGO events in Section V). The talk only gave a few references (with none to myself) and I do the same here. If there is a problem locating references, please contact me.}
\affiliation{Department of Physics, University of Toronto, Toronto, Ontario, Canada  M5S1A7}

\begin{abstract}
We encounter these demons on the path towards a UV complete QFT of gravity and a horizonless replacement for black holes. The fate of the ghost is discussed in the strong coupling version of classically-scale-invariant quadratic gravity. Rather than a propagating ghost, the full graviton propagator ends up with a slight acausal behavior. We then turn to the 2-2-hole solutions appearing in a classical approximation of the gravity theory. We present a new solution that is sourced by a relativistic gas, where a benign timelike singularity is shrouded by a fireball. Calculating the standard entropy gives an area law, and the entropy of a 2-2-hole can exceed that of the same size black hole. Finally, an observational consequence of 2-2-holes takes the form of gravitational wave echoes that can be generated by a newly formed 2-2-hole after a 2-2-hole merger. LIGO is sensitive to such a signal. We update the evidence from our previous search with five additional events reported more recently by LIGO.
\end{abstract}
\maketitle

\section{Introduction}

\begin{figure}[!h]
\centering
   \includegraphics[width=9cm]{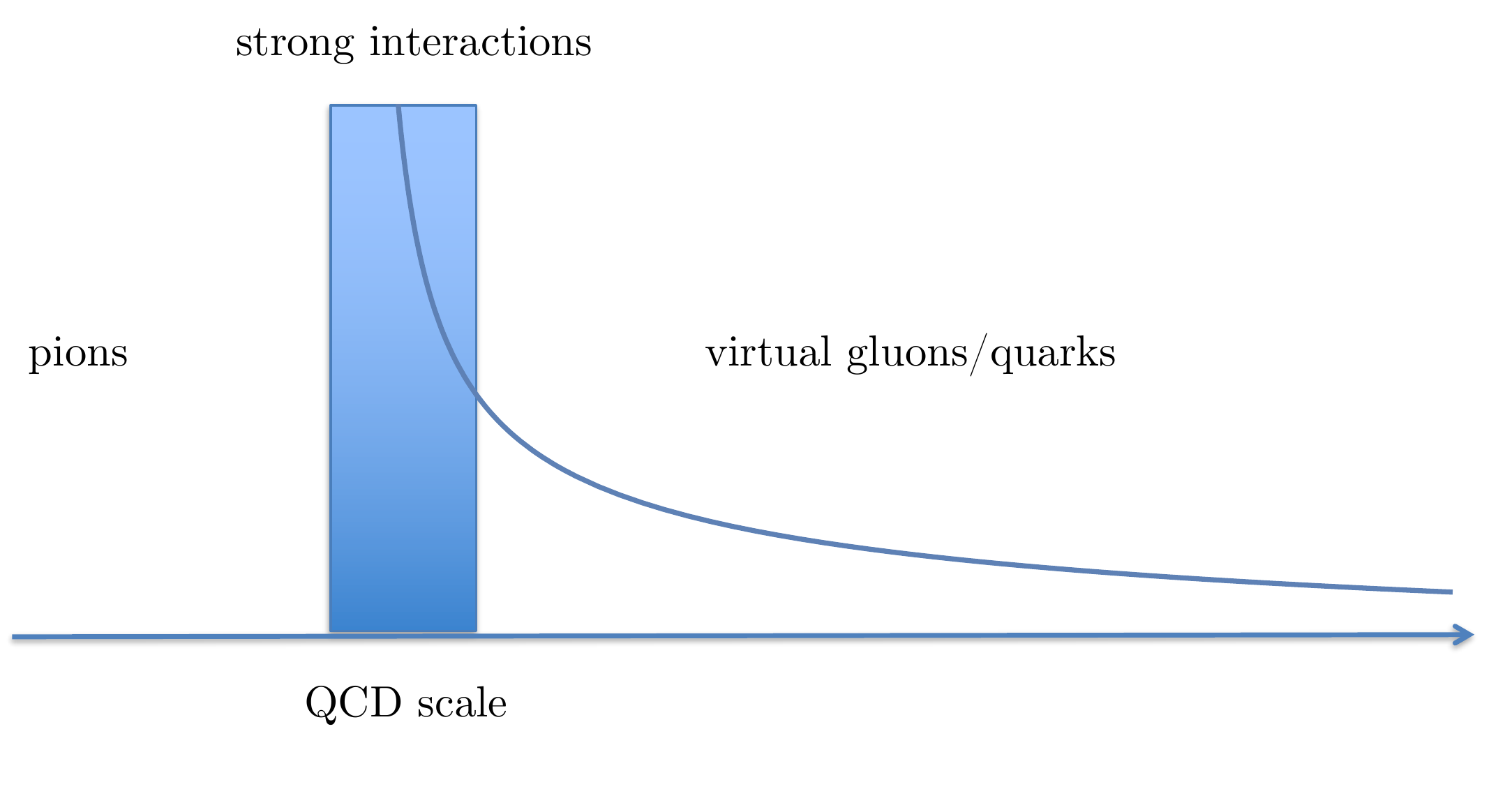} 
   \includegraphics[width=9cm]{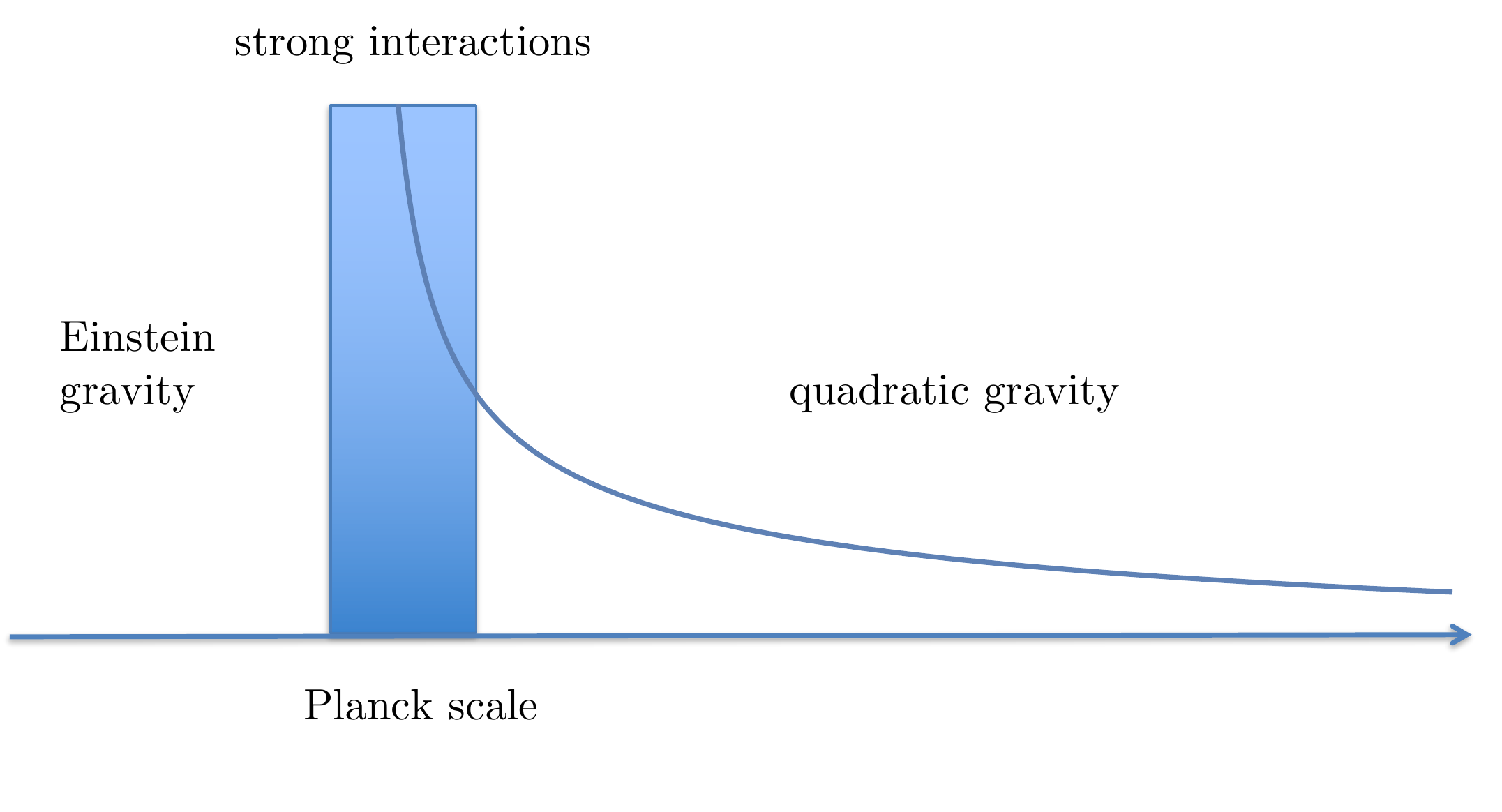}  
\caption{The proposed similarity between the dynamics at the QCD and Planck scales.}
\label{fig24}
\end{figure}

We start with Fig.~(\ref{fig24}a), which shows the known way that nature generates a mass scale dynamically. This is QCD. The central hypothesis of this talk is that the Planck mass scale is generated in a very analogous way, as shown in Fig.~(\ref{fig24}b). The running coupling that appears in Fig.~(\ref{fig24}b) is an asymptotically free coupling of quadratic gravity. This and the fact that quadratic gravity is renormalizable has been known for decades. Quadratic gravity differs from QCD in that there are two possible mass scales. One is the scale $\Lambda_{\rm QG}$ where the asymptotically free coupling grow strong. The other, much more well known, is the explicit (or spontaneously generated) mass $M$ in the Einstein term $M^2R$. If $M\gg\Lambda_{\rm QG}$ then the running coupling remains weak and gravity is weakly coupled. This is the standard choice, and in the case of the agravity scenario (Salvio and Strumia 2014) it is chosen for a good reason; if gravity is weak enough (exceedingly so) then the gravity contributions to the Higgs mass may be under control.

We are instead considering the strong gravity option, in particular with $M=0$. This leaves us with a Higgs puzzle and for this we go to the other extreme and assume that there are no elementary scalar fields in nature. In particular it is still not certain that the observed Higgs boson is an elementary particle. We are also drawn to the idea that the fundamental theory contains no mass parameters of any kind, and is thus classically scale invariant. This leads into some of our thoughts about the cosmological constant, but since this is a little removed from the present discussion, we place these thoughts in the Appendix.

\begin{figure}[!h]
\centering
   \includegraphics[width=9cm]{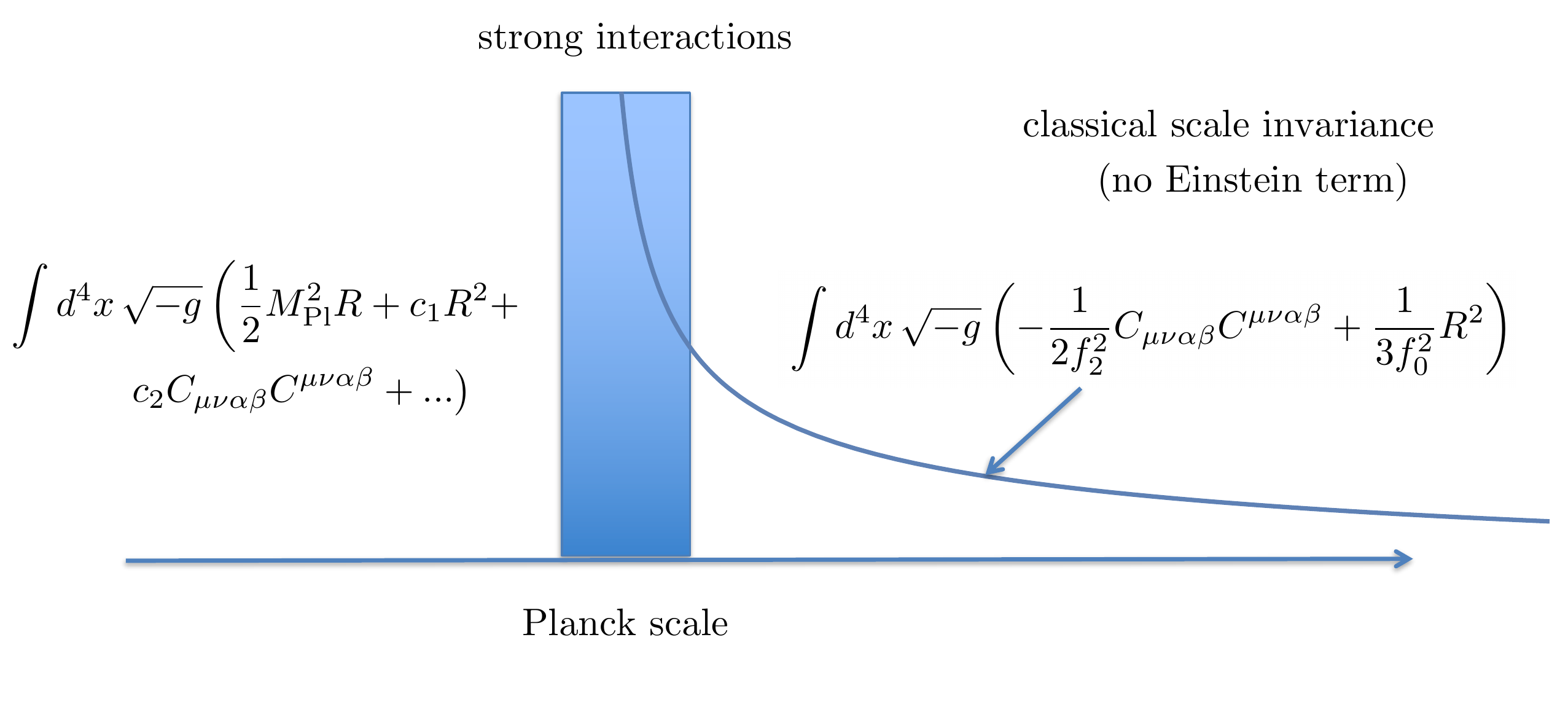} 
\caption{The fundamental theory above the Planck scale, and the effective theory below.}
\label{fig25}
\end{figure}

We turn to a more detailed version of Fig.~(\ref{fig24}b) as shown in Fig.~(\ref{fig25}).
The classically invariant version of quadratic gravity has two running couplings, with the $f_2$ appearing with the Weyl-curvature-squared term being asymptotically free. The scale at which the asymptotically free coupling $f_2$ grows strong defines a mass scale and this provides the dynamical origin of the Planck mass. $f_0$ is typically not asymptotically free, but Salvio and Strumia (2018) argue that the corresponding degrees of freedom decouple in the UV, thus not ruining the UV-completeness of the theory. The effective theory in the infrared is the standard derivative expansion that respects the symmetries and that involves the dynamically generated mass scale. We are assuming that the diff invariance is not broken by the strong dynamics, as is the case for gauge invariance in the QCD analogy. The Einstein term that appears in the effective theory and which dominates at low energies is not fundamental. A cosmological constant term is not included, and the Appendix describes our reasoning for that.

\section{Fate of ghost in strongly interacting quadratic gravity}
\label{s2}

It has been widely publicized that the theory of quadratic gravity suffers from some kind of ghost problem associated with its four-derivative structure. Our intuition concerning ghosts has been largely formed in the classical context or in toy quantum theories. Here we need to reconsider the issue in a theory that is more QCD-like, where the mass scale is generated dynamically. As already indicated the quantum theory is defined by the following action with classical scale invariance,
\begin{align}
S_{QG}=\int d^4x\,\sqrt{-g}\left(-\frac{1}{2 f_2^2}C_{\mu\nu\alpha\beta}C^{\mu\nu\alpha\beta}+\frac{1}{3 f_0^2}R^2\right).
\end{align}

\begin{figure}[!h]
\centering
   \includegraphics[width=9cm]{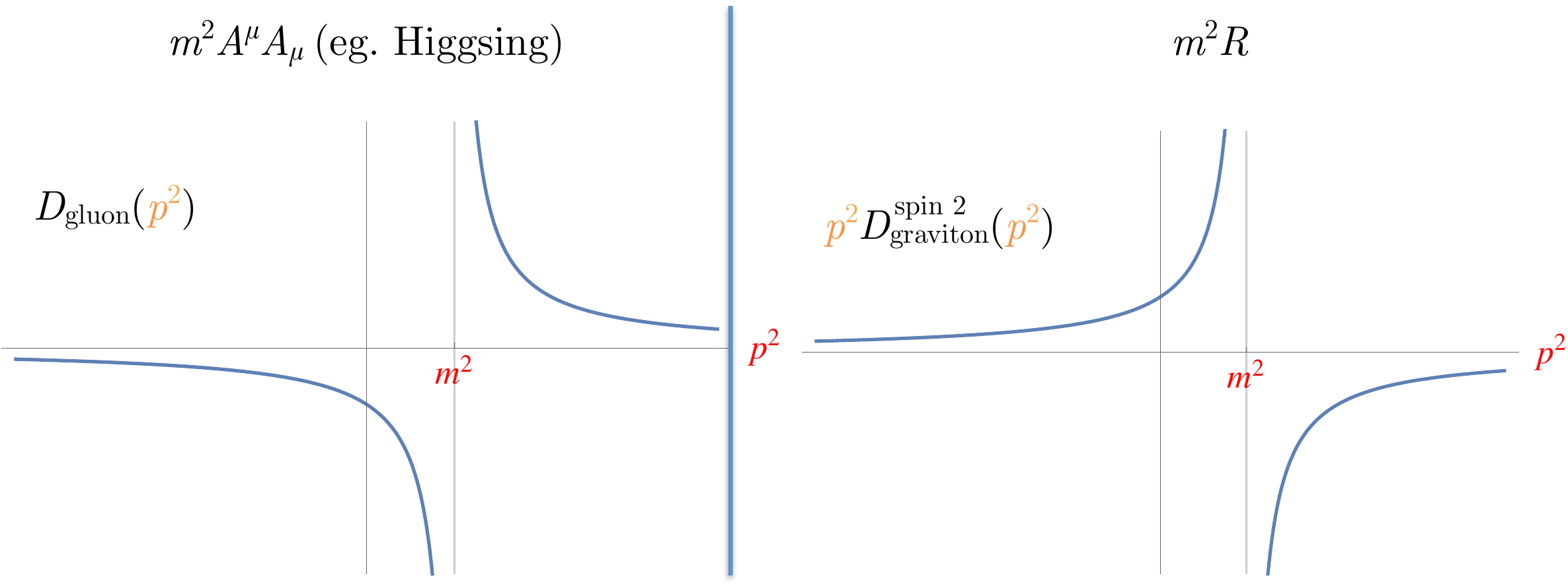}  
\caption{Comparing the tree level gluon and $p^2$ times the spin-2 graviton propagator after naive mass terms have been added. The main difference is an overall sign.}
\label{fig12}
\end{figure}

To discuss ghosts we need to discuss propagators, which are of the form $iD(p^2)\times \textrm{(tensor structure)}$. We shall display some plots of $D(p^2)$'s, showing the real part, and we start by comparing gluons and gravitons. We consider the tree level propagators after adding a naive mass term to each theory, as shown in Fig.~(\ref{fig12}). For QCD this could involve the Higgs mechanism to produce a $m^2A^\mu A_\mu$ term. For gravity it is simpler since the gauge (diff) invariant Einstein term is the mass term. In QCD we would be left with a propagating massive gluon, which we know does not exist, while in gravity we would be left with a propagating massive ghost.

\begin{figure}[!h]
\centering
   \includegraphics[width=9cm]{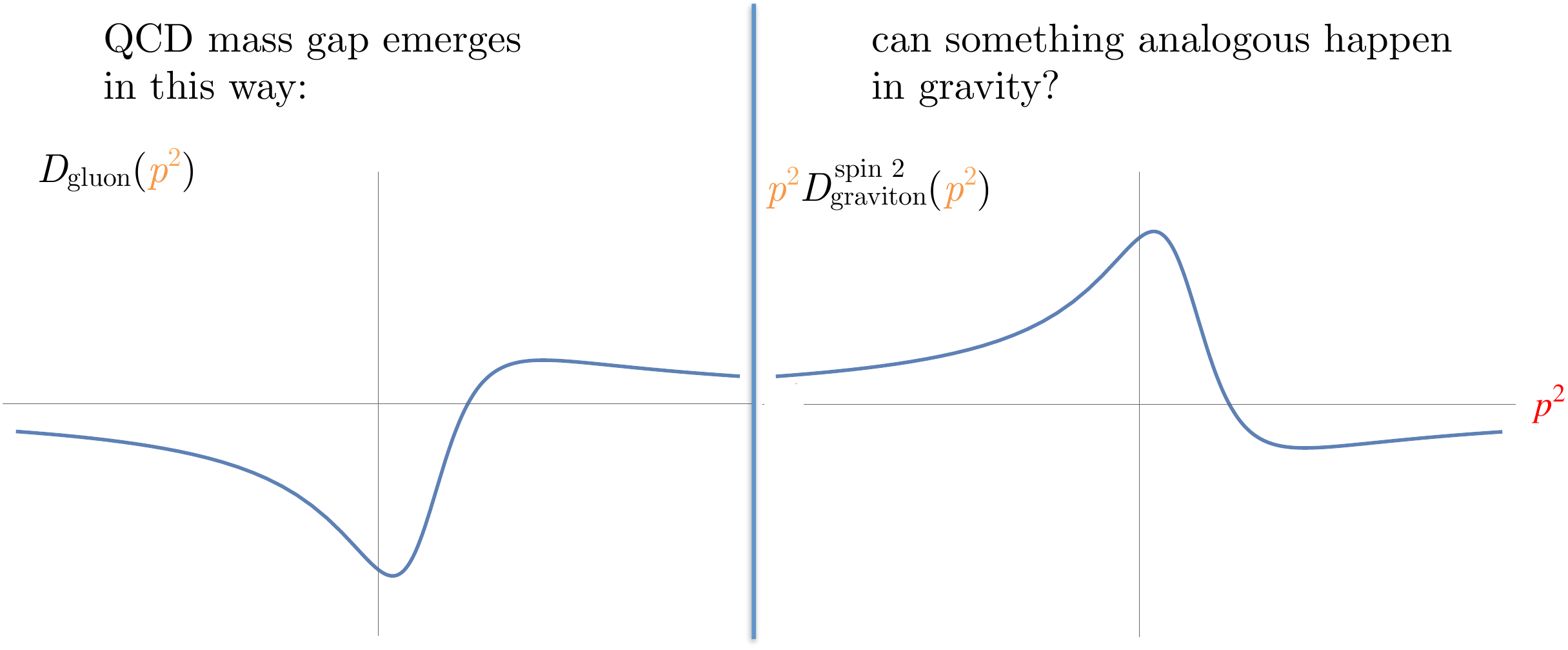}  
\caption{The realistic full gluon propagator and the inverted possibility for strong gravity.}
\label{fig13}
\end{figure}

Fig.~(\ref{fig13}) shows the actual realistic form of the full gluon propagator. QCD manages to generate a mass gap without breaking the gauge invariance. There is no pole and thus no propagating massive gluon. Can we expect that strong gravity can generate a mass gap in a similar way, up to the different overall sign?

To explore this further we consider the form of the full graviton propagator,
\begin{align}
p^2 D^\textrm{spin 2}_\textrm{graviton}(p^2)=\frac{1}{M(p^2)^2-(1+{a(p^2)})p^2+ib(p^2)p^2\theta(p^2)}.
\end{align}
We have added some new terms to the $-p^2$ tree-level term. Strong gravity generates a mass scale and so it is expected that the $M^2$ term appears. We also include the $a$ and $b$ terms that have perturbative contributions, but that can be of order one at strong coupling. The imaginary term $b$ is generated by all kinematically allowed decays of a virtual $p^2>0$ graviton to other particles. These particles may include massless gravitons, since although this amplitude vanishes at lowest order, it may not vanish at higher order. For $M$ we know that $M(0)^2>0$ corresponds to the correct sign of the Einstein term, and we expect that $M(p^2)^2\to0$ quickly for $|p^2|$ large. $b(p^2)>0$ corresponds to perturbative unitarity (Donoghue and Menezes 2018). $a(p^2)$ and $b(p^2)$ tend to zero more slowly for $|p^2|$ large by asymptotic freedom. A mass $m^2$, usually called a ghost mass, is defined by where the real part vanishes, $M^2(m^2)-m^2(1+a(m^2))=0$.

\begin{figure}[!h]
\centering
   \includegraphics[width=9cm]{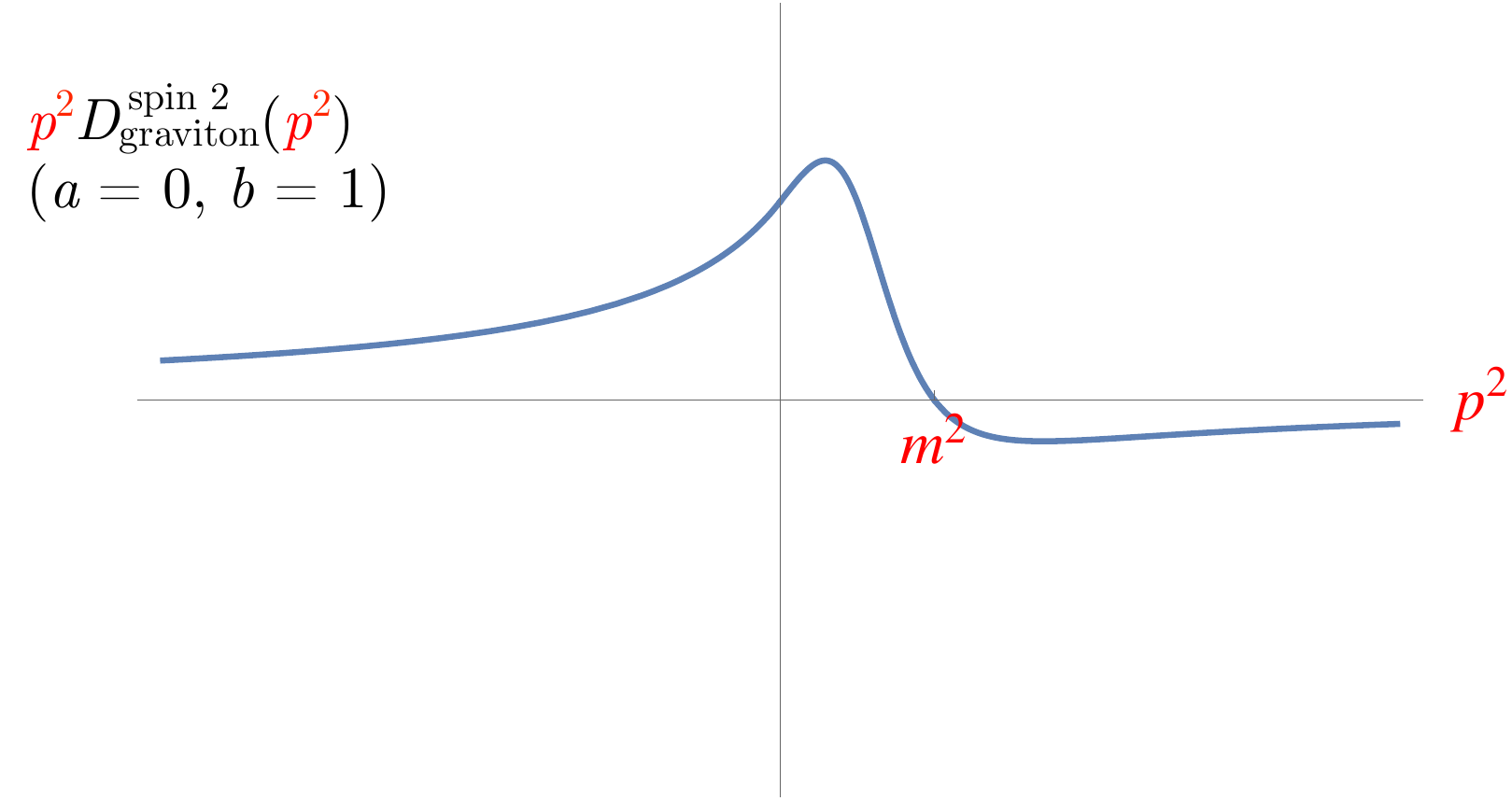}  
\caption{Real part of full graviton propagator.}
\label{fig14}
\end{figure}

Now if we consider the real part of this propagator, illustrated for some unimportant choice of $a$ and $b$, we get the result of Fig.~(\ref{fig14}). This form is what we guessed above in analogy with QCD. This basic shape has emerged from two phenomena: 1) a dynamical generation of a gauge invariant mass term and 2) a perturbative generation of an imaginary part. Both of these are natural and expected. On the other hand it is interesting that neither occur for the QCD gluon. There the less trivial confinement mechanism is at play. Thus one could say that the full graviton propagator is easier to understand than the full gluon propagator.

Let us consider an idealized version of the full propagator,
\begin{align}
D_\textrm{dressed}(p^2)=\frac{1}{p^2+i\epsilon}\;\frac{1}{-(p^2-m^2)+i b p^2}\quad\textrm{with }b>0.
\label{e1}
\end{align}
We would like to compare this ``dressed'' propagator to a bare propagator that was used by Stelle (1977) to demonstrate the renormalizability of the perturbative theory (in his case quadratic gravity with the Einstein term added explicitly). This requires an $i\epsilon$ term of the appropriate sign,
\begin{align}
D_\textrm{bare}(p^2)=\frac{1}{p^2+i\epsilon}\;\frac{1}{-(p^2-m^2)-i \epsilon}.
\end{align}
Let us compare these propagators of two different theories by transforming to coordinate space and considering a timelike separation,
\begin{align}
\bar D(t)=\int d^4p\,i\,D(p^2) e^{-i p^0 t}.
\end{align}
$\bar D(t)$ will be a dimensionless function of $m|t|$. In plots we will instead show another quantity $D_+(t)$ that contains only the positive energy contributions ($e^{-i E_\textbf{p}t}$, not $e^{i E_\textbf{p}t}$) and that has a prefactor that improves the look of the plots,
\begin{align}
D_+(t)\equiv -m|t| \bar D_\textrm{positive energy}(t).
\end{align}

\begin{figure}[!h]
\centering
   \includegraphics[width=7cm]{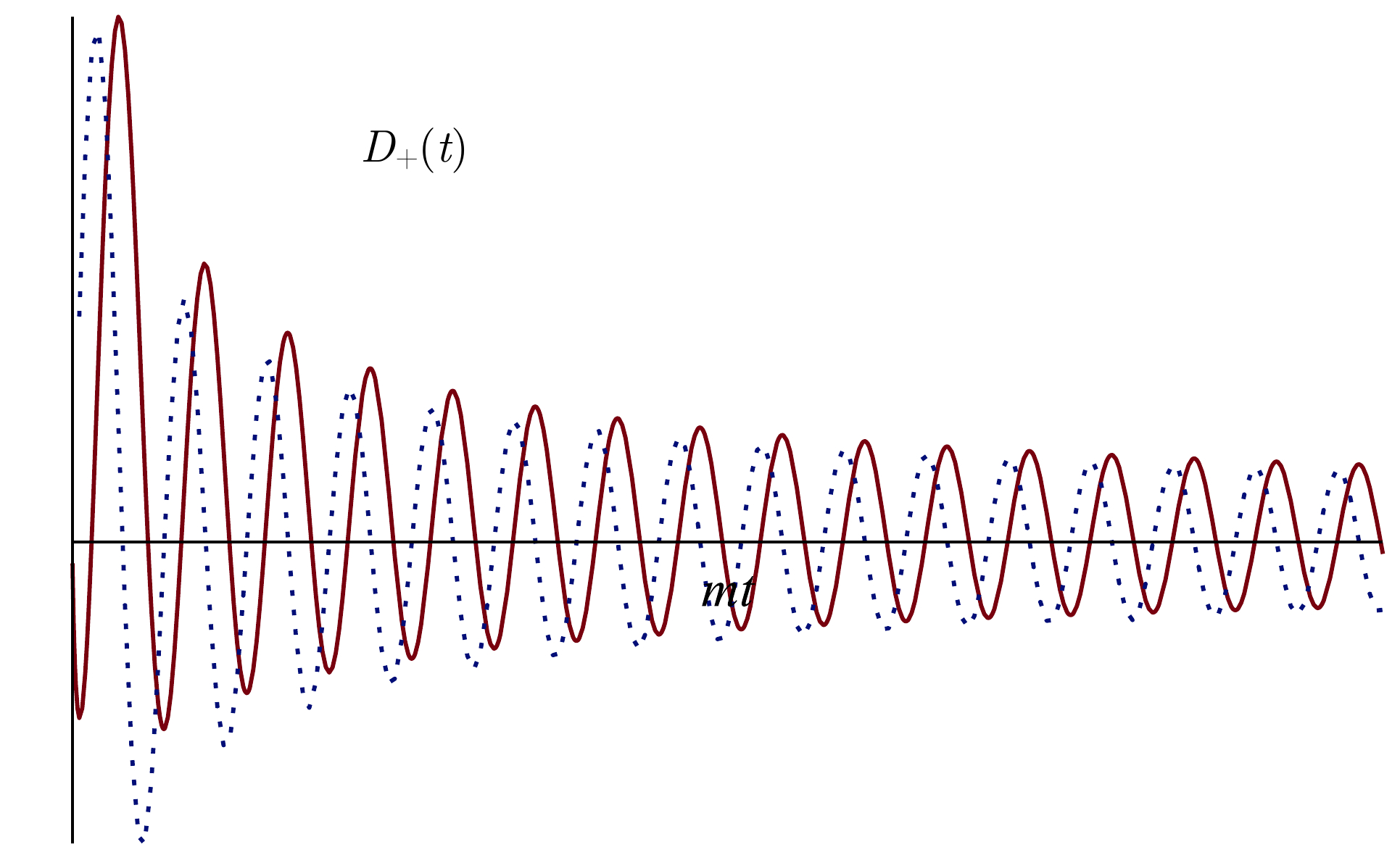}  
\caption{Real (solid) and imaginary (dashed) parts of $D_+(t)$ for the bare propagator.}
\label{fig15}
\end{figure}

\begin{figure}[!h]
\centering
   \includegraphics[width=7cm]{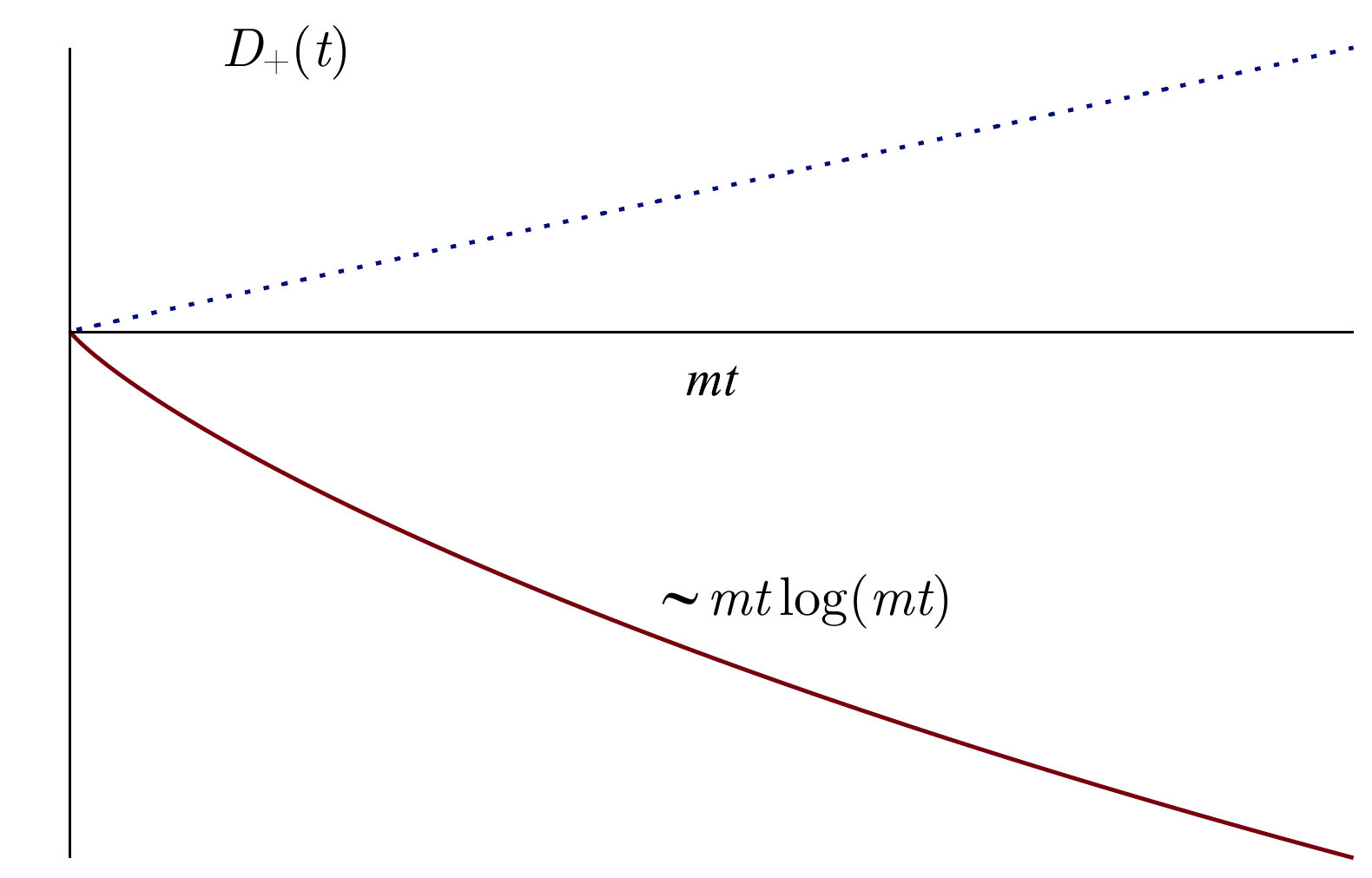}  
\caption{$D_+(t)$ for the bare propagator in the small $mt$ limit.}
\label{fig16}
\end{figure}

Looking at the bare propagator first we have
\begin{align}
\bar D(t)=-\frac{\pi}{m^2t^2}+\frac{i\pi}{m|t|} K_1(i m|t|),
\end{align}
where a Bessel function appears. The first and second terms correspond to the massless graviton and a propagating massive ghost respectively. The latter for large $t$ behaves like 
\begin{align}
\sim \frac{1}{(m|t|)^\frac{3}{2}}e^{-im|t|}.
\end{align}
$D_+(t)$ is shown in Fig.~(\ref{fig15}) where the oscillating ghost part dominates.
The only thing peculiar about the ghost term is its overall sign (corresponding to a negative-norm state), and this leads to a cancellation between the two terms at small $t$. Thus in the UV limit (the small $mt$ limit) we see from Fig.~(\ref{fig16}) that $D_+(t)$ tends to zero. This soft UV behaviour corresponds to the $1/p^4$ behavior at large $p$.

\begin{figure}[!h]
\centering
   \includegraphics[width=9cm]{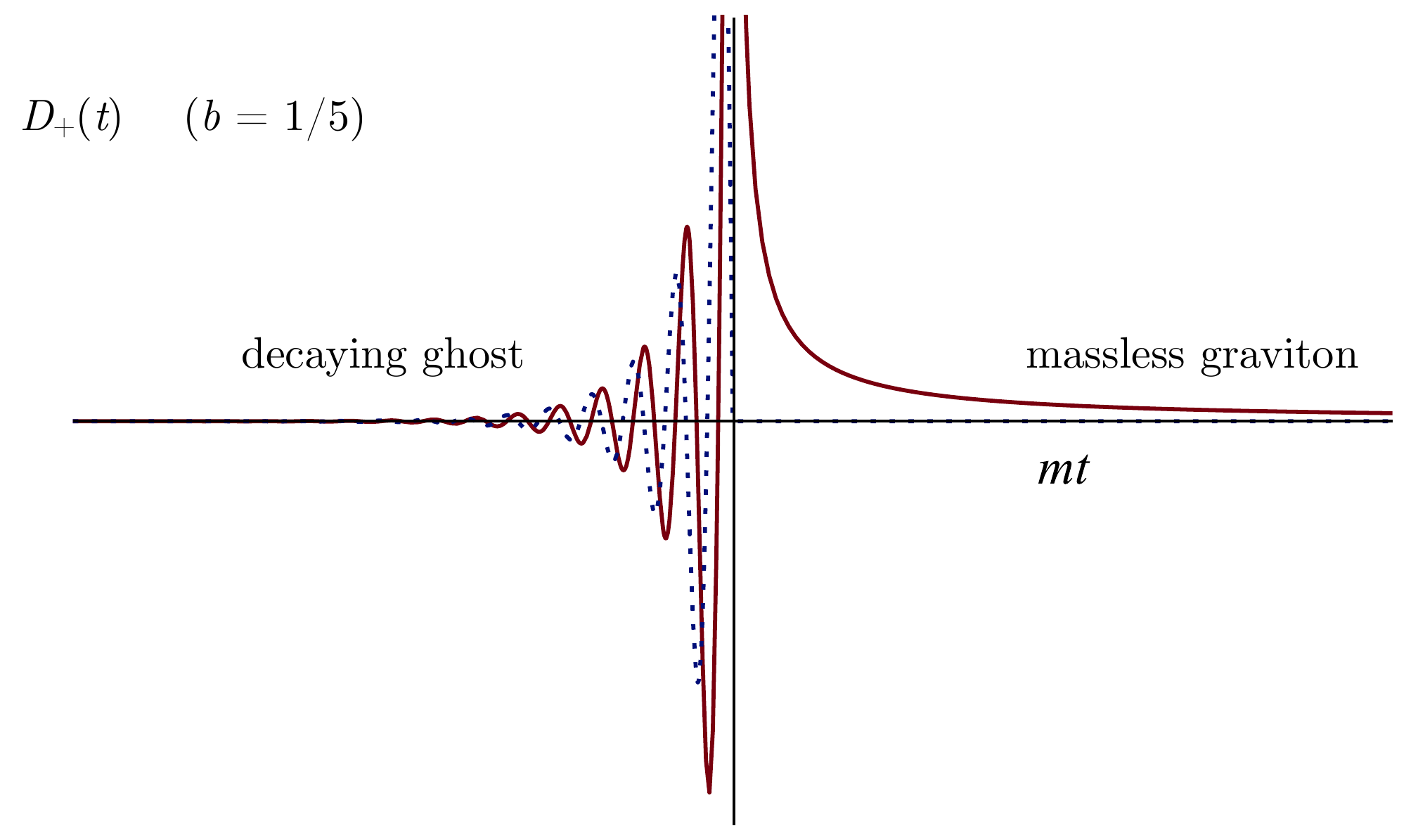}  
\caption{$D_+(t)$ for the dressed propagator.}
\label{fig17}
\end{figure}

The dressed propagator in (\ref{e1}) instead gives
\begin{align}
\bar D(t)=-\frac{\pi}{m^2t^2}+\frac{i\pi}{\sqrt{1-ib}}\frac{1}{m|t|} K_1(\frac{-i m|t|}{\sqrt{1-ib}}).
\end{align}
Now the large $t$ behaviour of the ghost term is
\begin{align}
\sim \frac{1}{(m|t|)^\frac{3}{2}}e^{+im|t|-\frac{1}{2}b|t|}.
\end{align}
There are two changes; the phase of the oscillatory factor has changed sign and a decay factor has appeared. The sign change means that positive (negative) energy now propagates backward (forward) in time. Fig.~(\ref{fig17}) shows $D_+(t)$ where we see a positive energy ghost excitation decaying into the past. Such an acausal amplitude can be probed for example in a scattering with the ghost in the $s$-channel, as discussed in detail by Grinstein, O’Connell and Wise (2008). Such behavior is nonstandard, but it appears not to be inconsistent.

\begin{figure}[!h]
\centering
   \includegraphics[width=9cm]{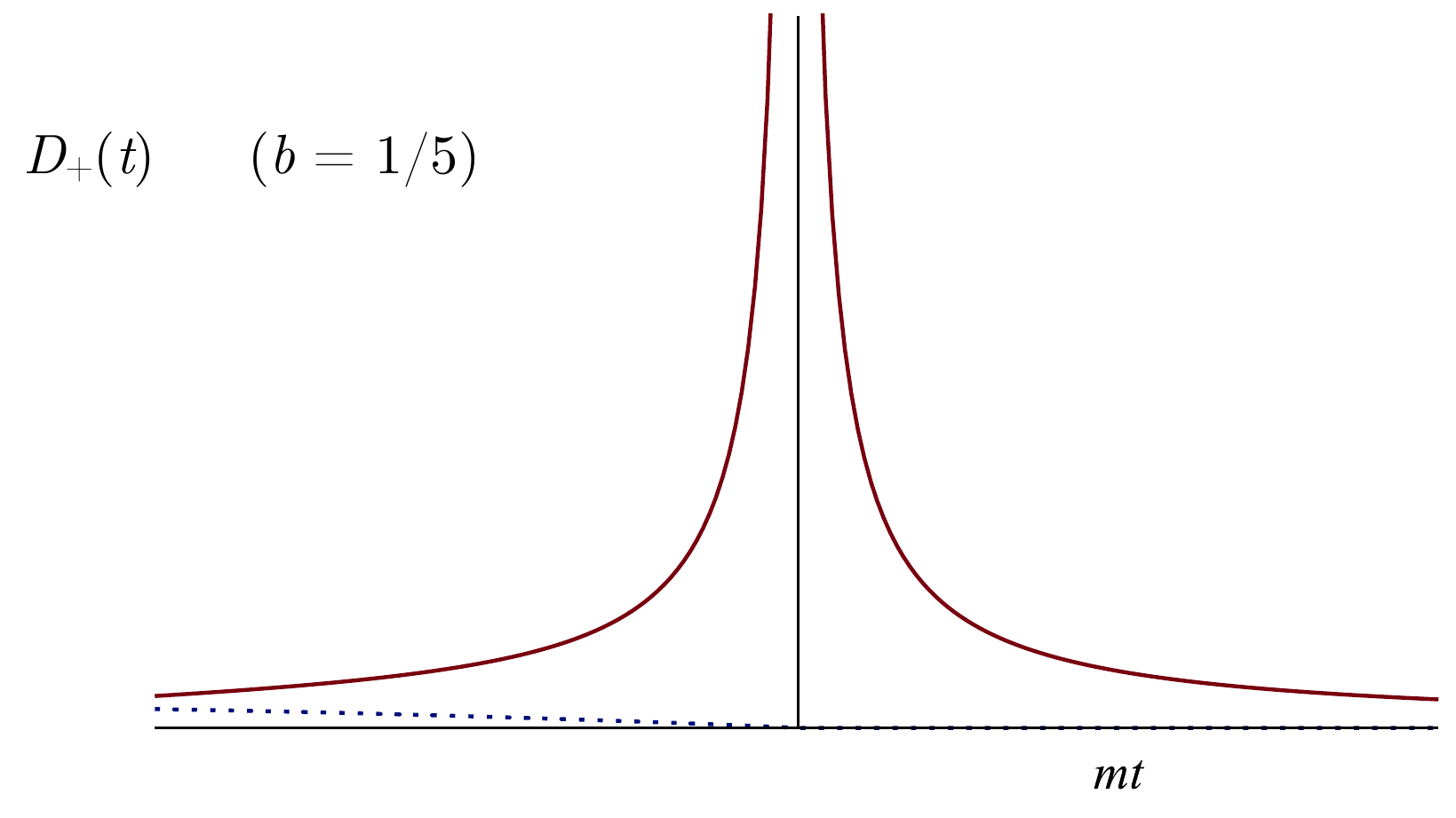}  
\caption{$D_+(t)$ for the dressed propagator in the small $mt$ limit.}
\label{fig18}
\end{figure}

Now if we consider the UV limit, the cancellation we had before no longer occurs. This is shown in Fig.~(\ref{fig18}). The time scales here are very short of order $t\sim 1/p_0$ for $p_0\gg m$, and so we are seeing a harder behavior at large $p_0$.
The very different behaviours of the bare and dressed propagators in the UV limit can be expressed as the difference of the two propagators in momentum space. We find
\begin{align}
[iD_\textrm{dressed}(p^2)-iD_\textrm{bare}(p^2)]\;\to\; 2\pi\delta(m^2p^2)\quad\textrm{for }p_0/m\to\infty.
\end{align}
Away from this limit the difference in the propagators is more complicated, as seen in the previous figures.

\begin{figure}[!h]
\centering
   \includegraphics[width=9cm]{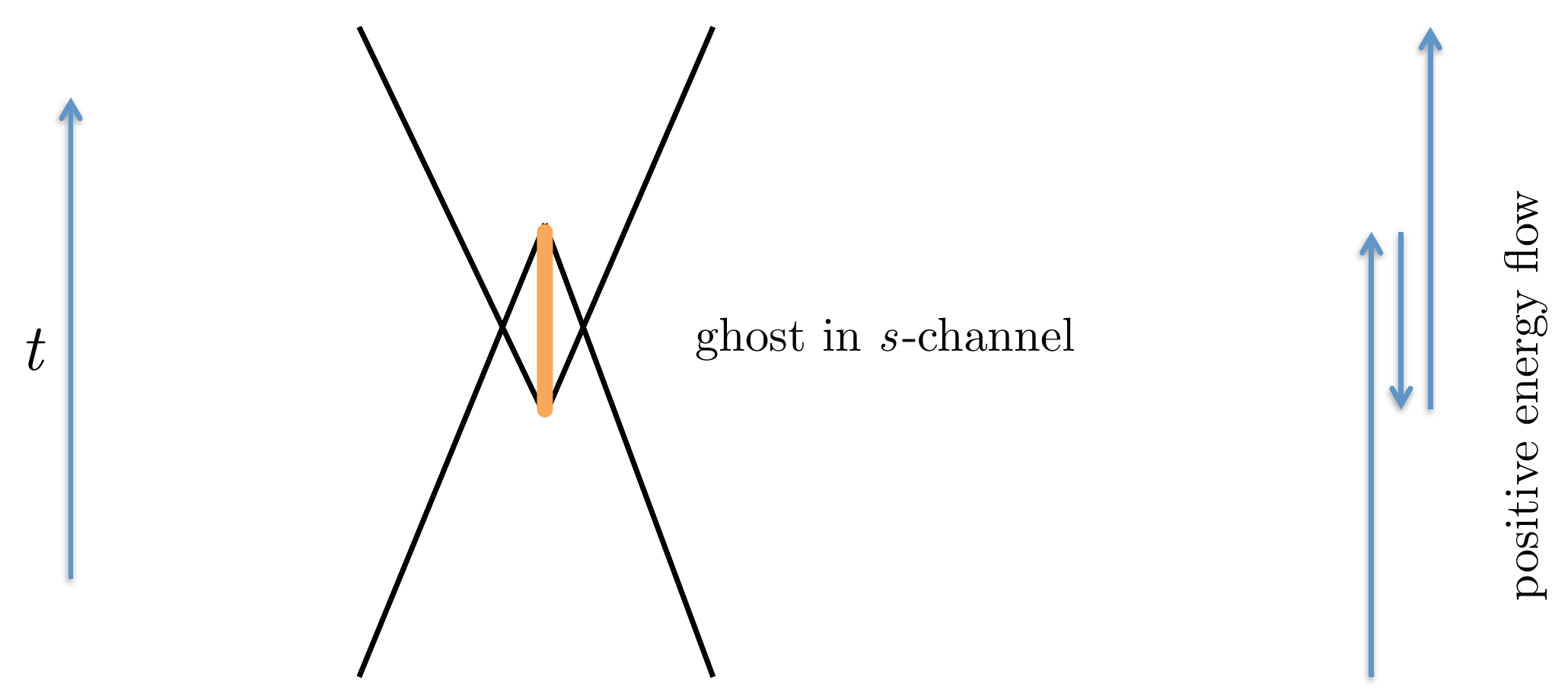}  
\caption{Decay products of the ghost emerge before the annihilation that produces the ghost.}
\label{fig19}
\end{figure}

The dressed propagator then indicates that the ghost only causes a short acausal transient of order the Planck time, $\Delta t\sim t_\textrm{Pl}/b$ (Fig.~\ref{fig17}). Since the Planck scale is where gravity becomes strong, we already expect large fluctuations of spacetime at these scales, and so the acausal behavior of the dressed propagator may be considered even less unusual. It could be said that there is a ``pre-disturbance'' to ghost production, from which the decay products of the ghost emerge. This is illustrated in Fig.~(\ref{fig19}), which also indicates the flow of positive energy. The final state only has normal particles, possibly including massless gravitons. This “deghostification” in gravity is analogous to hadronization in QCD.

To do perturbative calculations at super-Planckian energies, we would have to ensure that propagators have $p^2 \gg m_\textrm{Pl}^2$ to avoid strong coupling. In other words internal gravitons of a hard process should have high virtuality. For such calculations we should be able to use bare graviton propagators, in the same way that bare gluon propagators are used in perturbative QCD. In neither case are on-shell excitations considered.

The main conclusion here is that the full propagator does not describe a propagating ghost of negative norm. Instead the corresponding degree of freedom is now highly unstable while exhibiting a slight acausality upon its decay. Another feature of the full propagator is a dynamical mass that quickly fades for virtuality above the Planck scale.

\section{A black hole replacement and the fate of a singularity}

To study macroscopically large solutions of quadratic gravity we return to a classical approximation of the theory. The action for classical quadratic gravity (CQG) is
\begin{align}
S_{\mathrm{CQG}}=\frac{1}{16\pi}\int d^4x\,\sqrt{-g}\left(M_\textrm{Pl}^2R-\alpha C_{\mu\nu\alpha\beta}C^{\mu\nu\alpha\beta}+\beta R^2\right).
\end{align}
This action has high and low curvature limits similar to the quantum action, where quadratic terms dominate at high curvature, and the Einstein term (induced in the quantum case) dominates at low curvature. But the classical action interpolates between these two limits in its own way, and it also lacks running couplings. In addition, time evolution in this theory could suffer from the classical ghost instability. We thus restrict our study of the classical theory to static solutions.

We study solutions that have an inner region of high curvature and an outer region of low curvature, where each of these regions should be well described by the classical action. There is some transition region for a relatively small range of radii where the classical theory is interpolating between the two limiting behaviours. This is where corrections from the quantum theory could enter. In our previous work we have used numerical analysis plus scaling arguments to show that there are macroscopically large static solutions of this type that have no horizon.

QCD provides an analogy for these macroscopically large solutions in the form of quark matter, a speculated ground state of QCD. Quark matter configurations can extend up to neutron star size. The basic mechanism is that a dense enough accumulation of quarks drastically lowers the value of the chiral condensate. This then lowers the quark mass and thus the energy of the configuration, and this can more than compensate for the energy needed to change the value of the condensate. In quadratic gravity, our finding is that a dense enough accumulation of matter drastically lowers the volume of spacetime. The volume element is thus the analog of the condensate in quark matter. And then because of the much smaller volume, the increased pressure of matter can resist further collapse.

In the following we first review some of our previous results on 2-2-holes. We use the coordinate system where
\begin{align}
ds^2=-B(r)dt^2+A(r)dr^2+r^2d\theta^2+r^2\sin^2\theta d\phi^2.
\end{align}
At small $r$, both $A(r)$ and $B(r)$ behave like $\sim r^2 +O(r^4)$, which is the reason for the name of these objects. In fact the solutions we study here have the following expansion at the origin where $B(\infty)\equiv1$,
\begin{align}
&A(r)=\textcolor{red}{a_2} r^2+a_4 r^4+a_6 r^6+\cdots\\
&B(r)\propto r^2+\textcolor{red}{b_4} r^4+a_6 r^6+\cdots.
\end{align}
All coefficients are determined by the values of $a_2$ and $b_4$. This may be compared to the star-like solutions that have
\begin{align}
&A(r)=1+\textcolor{red}{a_2} r^2+a_4 r^4+\cdots\\
&B(r)\propto1+\textcolor{red}{b_2} r^2+b_4r^4+\cdots.
\end{align}

The vanishing of the metric at $r=0$ gives rise to a naked time-like singularity, our second apparent demon. The radius of the object can be set equal to the black hole radius, $r_h=2GM$, with $M$ the mass that parameterizes the external Schd solution. Macroscopically large means that $r_h\gg\ell_\textrm{Pl}$ where $G\equiv\ell_\textrm{Pl}^2$. The Schd solution holds down to radii just slightly larger than $r_h$, that is a distance from $r_h$ of order a Planck length or less. Inside this radius the solution transitions to the high curvature interior solution.

The interior curvatures behave as follows. For $r\lesssim r_h$, a curvature invariant $I(r)$ with mass dimension $d_I$ takes the form
\begin{align}
I(r)\ell^{d_I}_\textrm{Pl}= f_I(\frac{r}{r_h})\quad\textrm{with }f_I(1)\sim1,
\end{align}
This simple behaviour is related to the scale invariance of the quadratic terms in the action that determine the interior solution. In particular the Weyl curvature invariant behaves as the following for small $r$,
\begin{align}
C^2\ell_{\rm Pl}^4\sim \frac{r_h^4}{r^4}.
\end{align}
The volume element also displays simple scaling at small $r$,
\begin{align}
\frac{\sqrt{-g}}{\ell_{\rm Pl}^2}\sim \frac{r^4}{r_h^4}.
\end{align}
The latter means that in addition to the total mass $M$, various integrals, such as the action, remain finite. The curvature scalar $R$ itself is finite and so the $C^2$ term dominates the $R^2$ term in the action in the interior.

The metric functions have a small prefactor at small $r$,
\begin{align}
A(r),B(r)\sim\textcolor{red}{\frac{\ell_{\rm Pl}^2}{r_h^2}}\;\frac{r^2}{r_h^2}.
\end{align}
Thus as we scale $r_h$ up to macroscopically large values, $A(r)$ and $B(r)$ become extremely tiny in the interior -- e.g. $\ell_{\rm Pl}^2/r_h^2\sim 10^{-80}$. This of course translates into a greatly reduced volume element. The small volume remaining inside $r=r_h$ is concentrated close to $r_h$. The ratio of $A(r)$ and $B(r)$ determines the apparent radial speed of light
\begin{align}
0<\frac{dr}{dt}=\sqrt{\frac{B(r)}{A(r)}}\leq1.
\label{e4}\end{align}
The speed is finite, even at the origin, and, as we shall see, it is smallest near $r=r_h$.

Consideration of the scalar wave equation will give some indication of the behavior of quantum test particles around the singularity. With $\Box\varphi=0$ where $\varphi=\sum_{lm}\psi_l(r,t)Y_{lm}(\theta,\phi)$ we have
\begin{align}
\partial _t^2 \psi_l =\frac{B}{A}\partial _r^2\psi _l+\frac{B}{A}\left(\frac{2}{r}+\frac{B'}{2B}-\frac{A'}{2A}\right)\partial _r\psi _l- B\frac{l (l+1)}{r^2}\psi _l.
\end{align}
The factor $B/A$ is the ratio we have just mentioned and the terms $B'/2B-A'/2A\sim r/r_h$ are sub-dominant to the $2/r$ term. The main change is in the last term, which is now finite at the origin as well as being suppressed by the smallness of $B$. We can compare the solutions of this equation to those for regular or flat spacetime. The behavior near the origin ($l=0,1,2,\dots$) is indicated in the table.
\begin{table}[h]
\begin{center}
\vspace{1em}
\begin{tabular}{|c|c|c|c|c|}
\hline
&&&&
\\[-3mm]
spacetime\,\, & $A(r)$ & $B(r)$ & $\psi_{l1}(r,t)$ & $\psi_{l2}(r,t)$
\\
&&&&
\\[-3.5mm]
\hline
regular & $r^0$ & $r^0$ & $r^{l}$ & $r^{-(l+1)}$
\\ 
\hline
2-2-hole & $r^2$ & $r^2$  & 1 & $r^{-1}$
\\
\hline
\end{tabular}
\end{center}
\end{table}
In both cases, just one solution is physical, i.e.~has finite energy in a finite volume that encloses $r=0$. The main difference for a 2-2-hole is that all waves behave as $l=0$ waves ($s$-waves) near the origin. The centrifugal barrier near the origin is missing. 

Because only one solution is physical, as in flat space, the singularity introduces no ambiguity or ill-defined time evolution for waves. This is how the time-like singularity can appear to be ``nonsingular'' when probed with quantum test particles. The fact that this can occur for certain types of timelike naked singularities is known in the literature. For us it then allows the study of hot quantum fields in the background of the 2-2-hole, in analogy with the brick wall model (t'Hooft 1985). The latter is a free field study, and for us we note that the high temperatures in the interior will probe the asymptotically free regime of the theory; that is the temperature reaches super-Planckian values in the interior. The fact that the corresponding wavelengths are small also means that the quantum field description may be replaced by a relativistic particle description (a perfect fluid with $\rho=3p$) to get the same results (Mukohyama and Israel 1998). But to be consistent with the quantum field description, the particle picture must be such that particles are not absorbed or emitted by the singularity.

\begin{figure}[!h]
\centering
   \includegraphics[width=9cm]{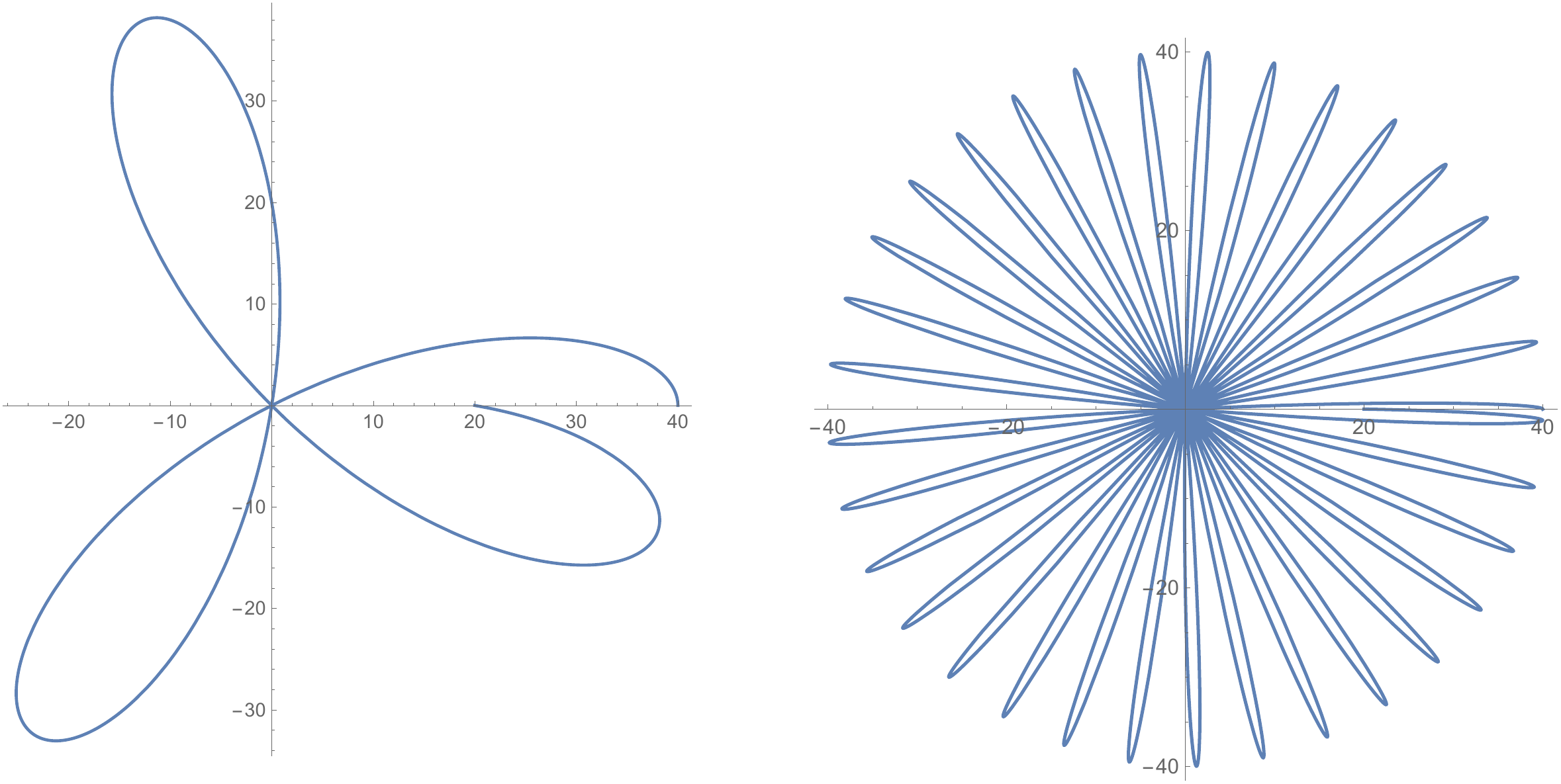}  
\caption{Examples of geodesics for a massive particle with large angular momentum $L$. $L$ is ten times smaller on the right.}
\label{fig11}
\end{figure}

We may digress further on the topic of the particle picture and point out that the geodesic equations suggest how to continue the motion of a particle through the singularity. In the region where $A(r)=ar^2$ and $B(r)=br^2$ is a good approximation, we can solve the geodesic equations for a massive particle to obtain
\begin{align}
r(t)=\sqrt{\frac{E^2}{b}-L^2}\;\cos(\frac{b}{E\sqrt{a}}t),\quad\quad\theta(t)=\frac{Lb}{E}t.
\end{align}
$E$ and $L$ is the particle energy and angular momentum divided by the particle mass. Note that when $r<0$ is encountered in polar coordinates, it is conventional for $(r,\theta)$ to mean $(-r,\theta+\pi)$. This then gives rise to smooth geodesics as a particle travels through the singularity. Examples of orbits are given in Fig.~(\ref{fig11}). We mention again that the classical point particle description is not the primary description. Any ambiguity in the particle description needs to be resolved to agree with the absence of absorption or emission by the singularity in the QFT picture, and we see that the geodesic equations point to a resolution that conserves $E$ and $L$ of the particle. 

From the explicit geodesic solution we see that $(L/E)_\textrm{max}=b^{-1/2}\sim r_h^2/\ell_\textrm{Pl}$ is very large while $\dot{\theta}_\textrm{max}=b^{1/2}$ is very small. For the largest $L$'s, the radial and angular speeds can be comparable, while for smaller $L$ the motion is mostly in the radial direction. Compared to the large range of possible $L/E$, we note that only particles with $(L/E)_\textrm{escape}\lesssim r_h$ can escape the 2-2-hole. ($E=1$ corresponds to the massive particle at rest at $r\to\infty$; thus massive particles also need $E>1$ to escape). This small fraction of particles with small enough $L/E$ to escape translates into an ``escape cone'' with a very small solid angle $\sim(\ell_\textrm{Pl}/r_h)^2$.

A 2-2-hole can also be considered as a particle collider. Two particles of mass $m$ dropped in and colliding in the interior can have super-Planckian center-of-mass energies,
\begin{align}
E_\textrm{CM}\sim \frac{2m}{\sqrt{B(r)}}\sim 2m\textcolor{red}{\frac{r_h}{\ell_\textrm{Pl}}}\frac{r_h}{r}.
\end{align}
Thus while the velocities of particles are subluminal, the components of a typical 4-momentum $p^\mu$ in the interior are enormous.

\section{An new 2-2-hole solution}
\label{s4}

Given this review of some our previous results, we now present a new result. We have found 2-2-hole solutions sourced by a relativistic perfect fluid gas ($\rho=3p$). These were found by starting the numerical integration close to the origin and using a high order series expansion to set the initial condition. Then the solutions were found by a shooting method with the goal of finding a Schd solution for $r>r_h$. Note that the $C^2$ term in the action is essential for these solutions while the $R^2$ term is optional. Thus studying these solutions becomes more tractable by considering the $\beta=0$ theory, that is a $R+C^2$ theory, since this reduces by one the number of parameters in the series expansion that need to be adjusted. (It also implies that $R=0$.) This theory still has a parameter that can be taken to be $m_2/m_\textrm{Pl}$, where $m_2$ is the naive ghost mass. Due to our picture of quantum gravity we expect this ratio to be of order unity, but we briefly keep it explicit in the following. 

These perfect fluid solutions have no analogs in general relativity, and they have some immediate physical implications. Given that the backreaction of the gas on the metric is fully accounted for, the thermodynamics of the gas and the geometry are explicitly linked. The thermodynamics and the particle content of the universe are also linked. Entropy arises as the standard entropy of ordinary hot quantum fields. An area law for entropy also arises.

The main results from this solution are as follows. The $r$ dependent temperature is $T(r)=T_\infty B(r)^{-1/2}$, as a simple consequence of hydrostatic equilibrium. The field equations then determine $A(r)$, $B(r)$ and the combination $NT_\infty^4$ where $N=\textrm{species number}\times m_2^2/m_\textrm{Pl}^2$. $\textrm{Species number}=1$ corresponds to one boson.
\begin{align}
&T_\infty\approx 1.33N^{-\frac{1}{4}}\textcolor{red}{T_\textrm{Hawking}}\\
&S=\frac{(2\pi)^3}{45}N\int_0^L T(r)^3A(r)^{1/2}r^2dr\approx0.75N^\frac{1}{4}\textcolor{red}{\frac{\textrm{Area}}{4\ell_\textrm{Pl}^2}}\label{e2}\\
&T_\infty S=\textcolor{red}{\frac{M}{2}}\\
&U=\frac{3}{4}T_\infty S=\frac{3}{8}M
\end{align}
The quantities in red are the black hole values. The entropy integral in (\ref{e2}) is finite even though the integrand diverges at $r=0$. (The upper limit on the integral is $L>r_h$ and we ignore the exterior volume $L^3$ contribution.) Thus while $T_\infty$ and $S$ differ from black hole values by a factor, the product $T_\infty S$ takes the black hole value. That product also defines the total thermal energy $U$, and its value happens to agree with the brick wall model in which a wall position is tuned ('t Hooft 1985).

One of the main observations here is that for a realistic number of species, and with the expectation that $m_2\approx m_\textrm{Pl}$, we find that the entropy of a 2-2-hole is greater than the same mass black hole. Given that our classical theory also has a BH solution, an entropy argument then indicates that a 2-2-hole rather than a BH is the favoured endpoint of a gravitational collapse. This is only a partial collapse, since matter remains distributed in the interior of a 2-2-hole rather than collapsing to a singularity. The small volume is related to high pressure of matter, and so in this way it is the small volume that resists collapse. A 2-2-hole has no maximum mass, but it does have a minimum mass (see below).

We see that the singularity is shrouded in its own fireball. Particles falling in are blue shifted to high energies, but they shower as they plow through the thermal bath, and eventually they equilibrate with it. The ghost of the quantum theory only causes a slight acausality in the steady state annihilation and production of normal particles, as described in Section II.

Standard quantum corrections to the vacuum energy are induced by the static curved background and this leads to the consideration of the Boulware vacuum state. For a black hole this state has a negative vacuum energy density that goes approximately like $1/B(r)^2$ outside the horizon (thus giving a diverging total energy). A Boulware vacuum state is the natural vacuum state for a static 2-2-hole spacetime, and the negative energy density can be expected to have the same $1/B(r)^2$ behavior. In this case the total negative energy is finite, and it can be largely compensated by the positive energy of a thermal gas at the appropriate temperature. This is referred to as a ``topped-up'' Boulware state, where by construction it causes minor backreaction (Mukohyama and Israel 1998). Thus the thermal excitation described above that sources the 2-2-hole is in addition to the thermal excitation of the topped-up state. Thus consideration of quantum corrections suggests that the total thermal energy and entropy and the actual $T_\infty$ are all larger than what is described above. We ignore this in the following.

\begin{figure}[!h]
\centering
   \includegraphics[width=9cm]{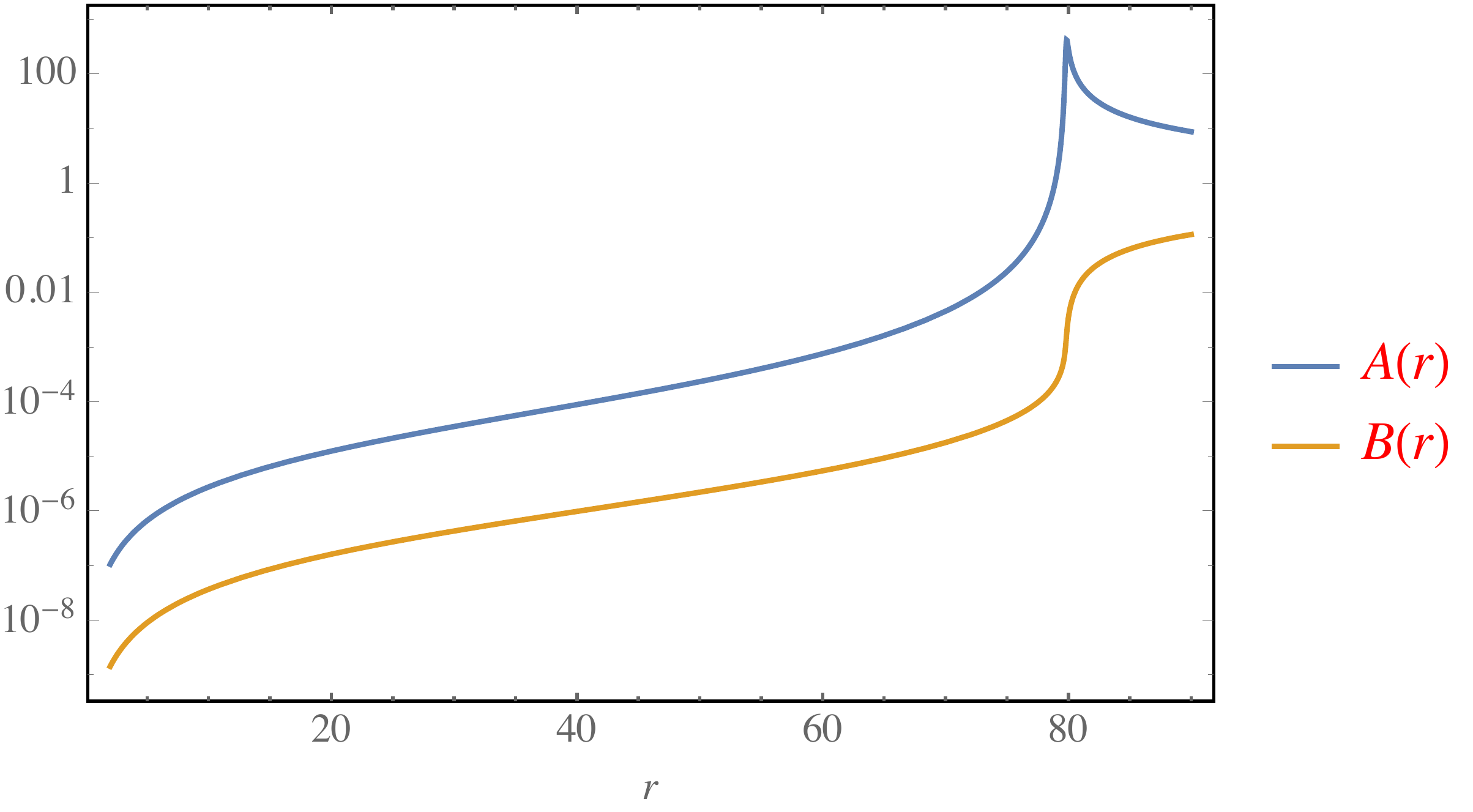}  
\caption{The two metric functions for $r_h=80\ell_\textrm{Pl}$.}
\label{fig1}
\end{figure}

\begin{figure}[!h]
\centering
   \includegraphics[width=9cm]{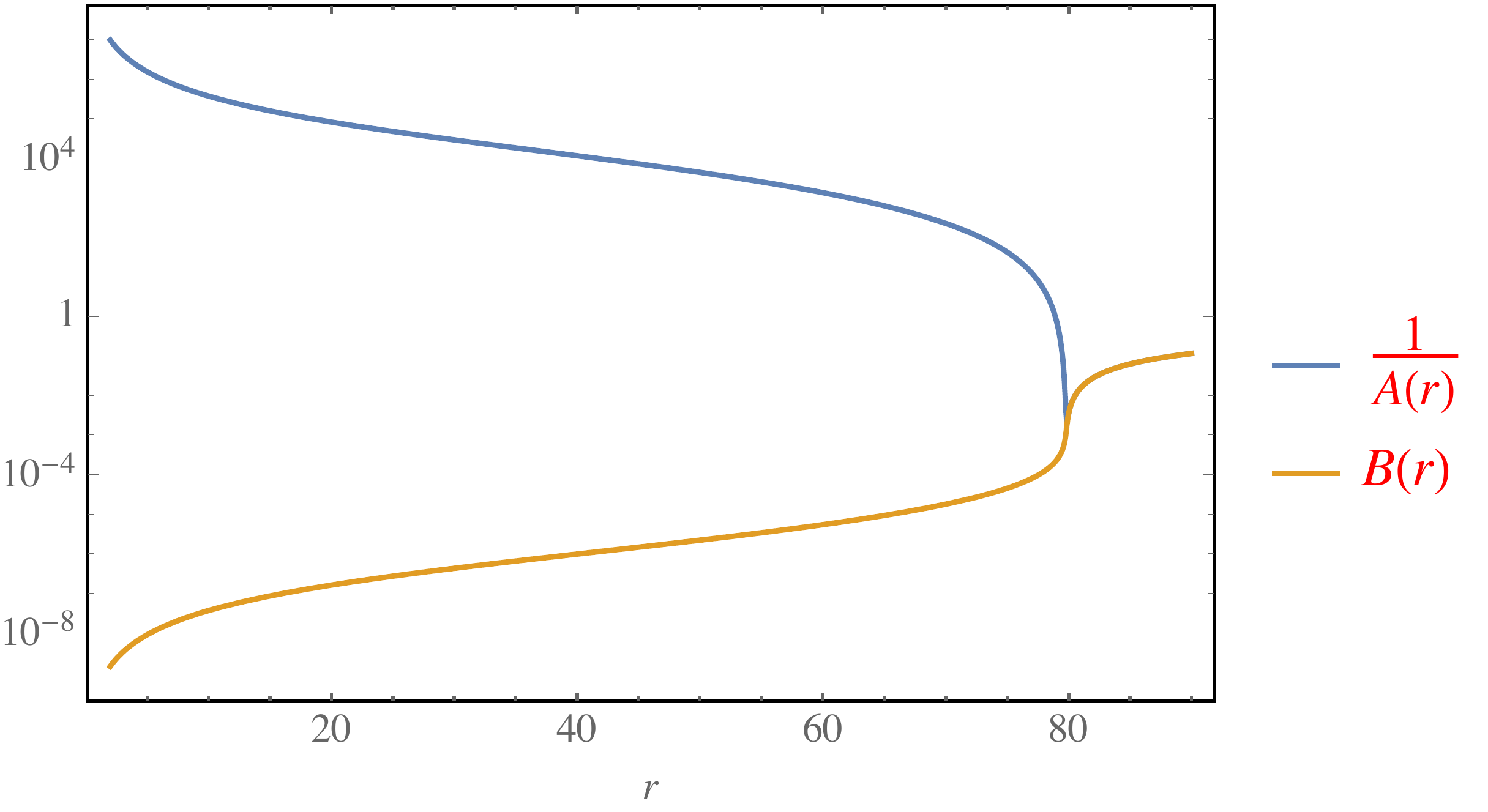}  
\caption{Inverting $A(r)$ more clearly shows the Schd behavior for $r>r_h$.}
\label{fig2}
\end{figure}

\begin{figure}[!h]
\centering
   \includegraphics[width=9cm]{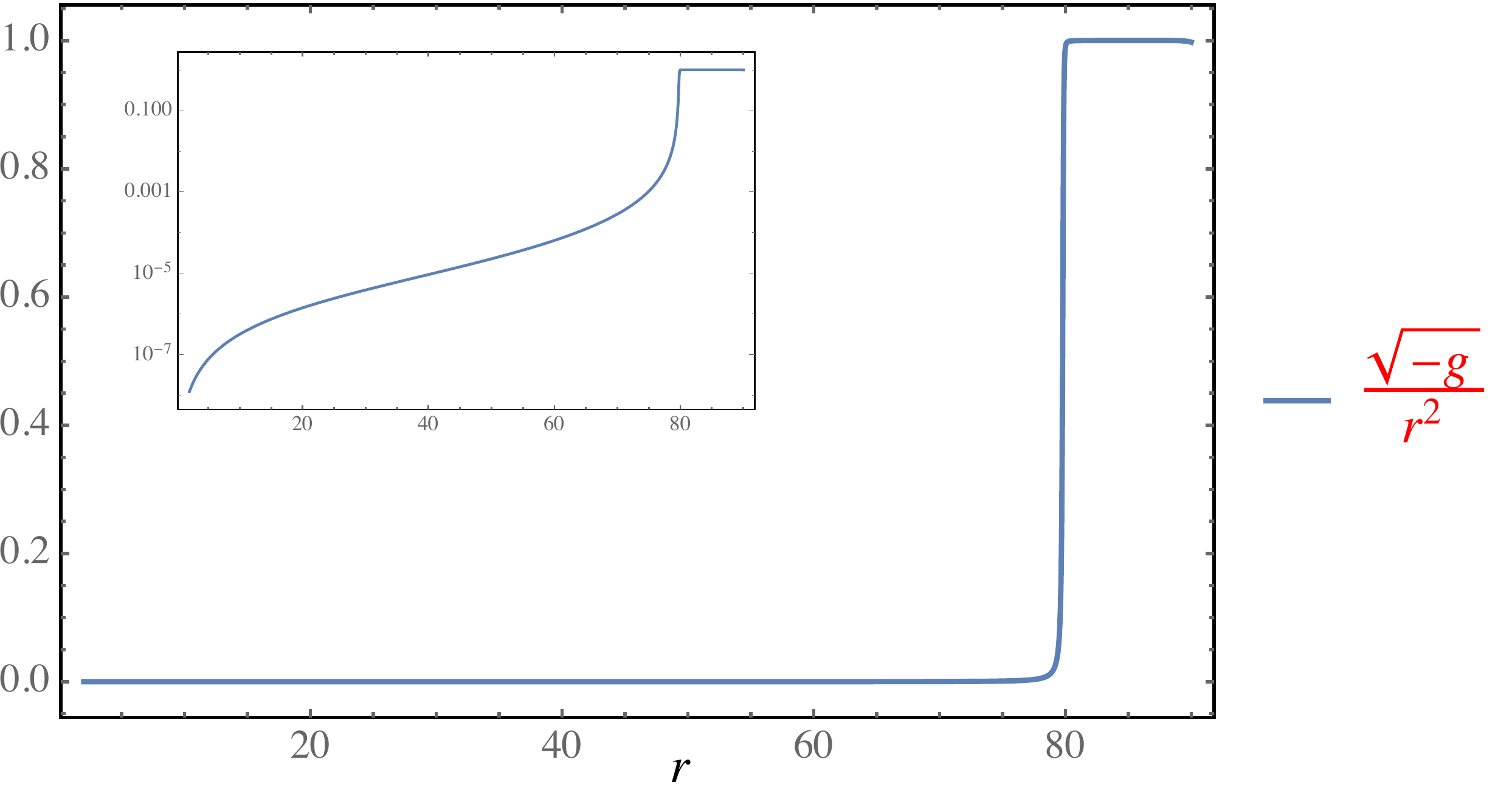}  
\caption{The volume element is drastically reduced in the interior, as shown on a linear and log scale (insert).}
\label{fig3}
\end{figure}

\begin{figure}[!h]
\centering
   \includegraphics[width=9cm]{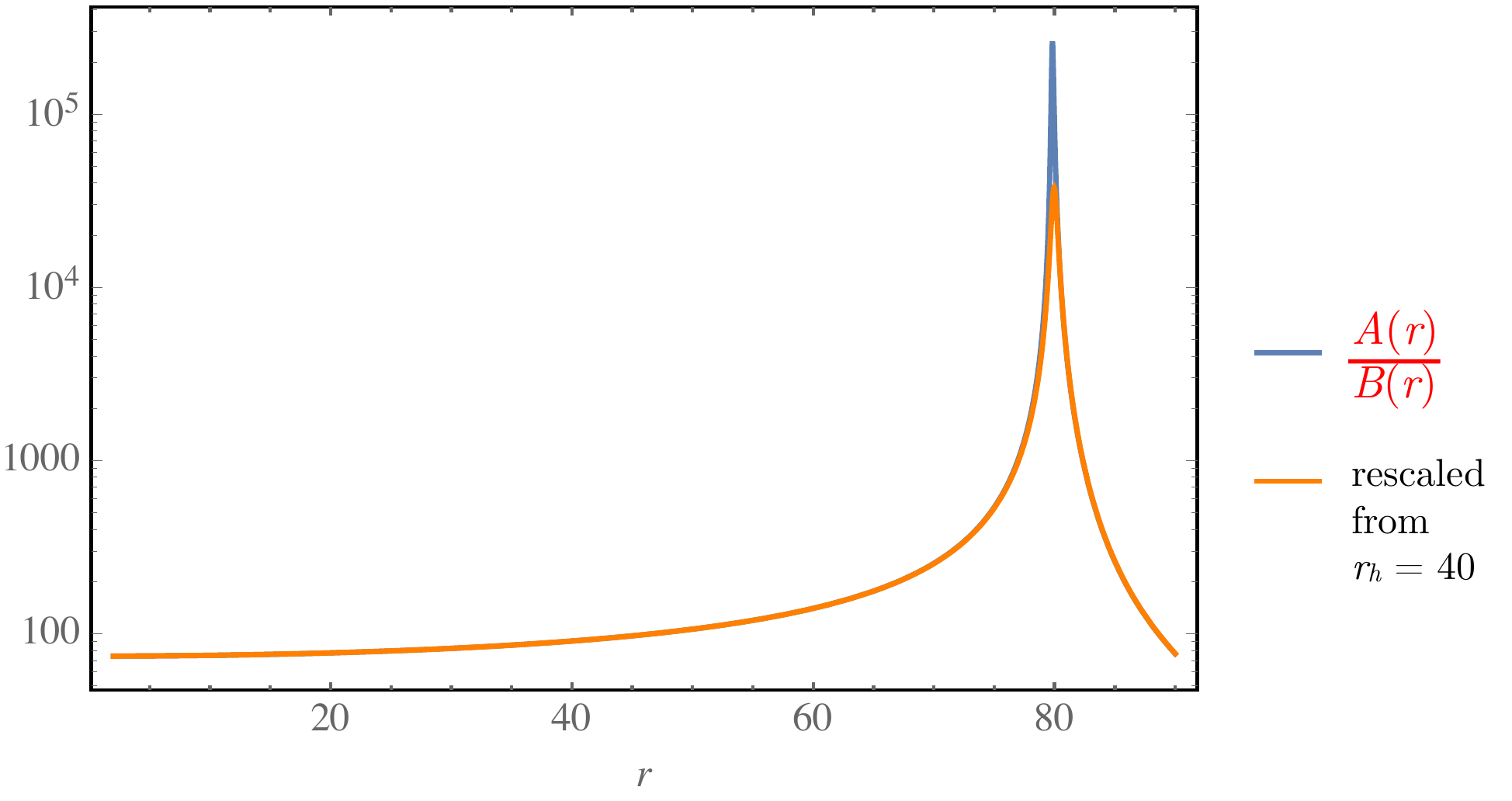}  
\caption{$A(r)/B(r)$ comparing $r_h=80$ and the rescaled $r_h=40$ solution. We see universal behavior everywhere except near $r=r_h$, where the peak height keeps growing for increasing $r_h$.}
\label{fig4}
\end{figure}

\begin{figure}[!h]
\centering
   \includegraphics[width=9cm]{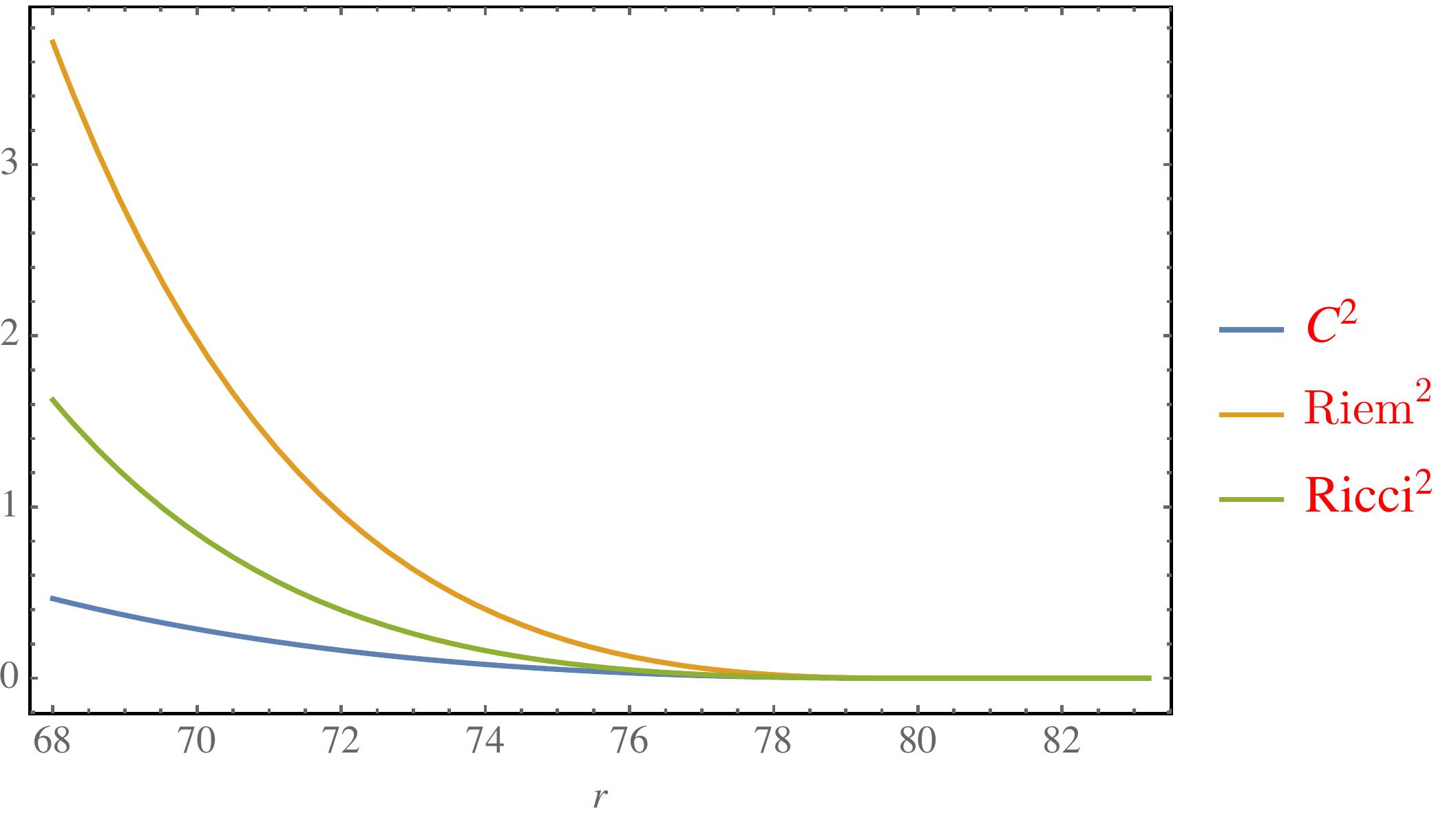}  
\caption{The dramatic feature in $A(r)/B(r)$ does not translate into a dramatic feature in the curvatures. They turn on smoothly, but note that this is a linear scale.}
\label{fig5}
\end{figure}

\begin{figure}[!h]
\centering
   \includegraphics[width=9cm]{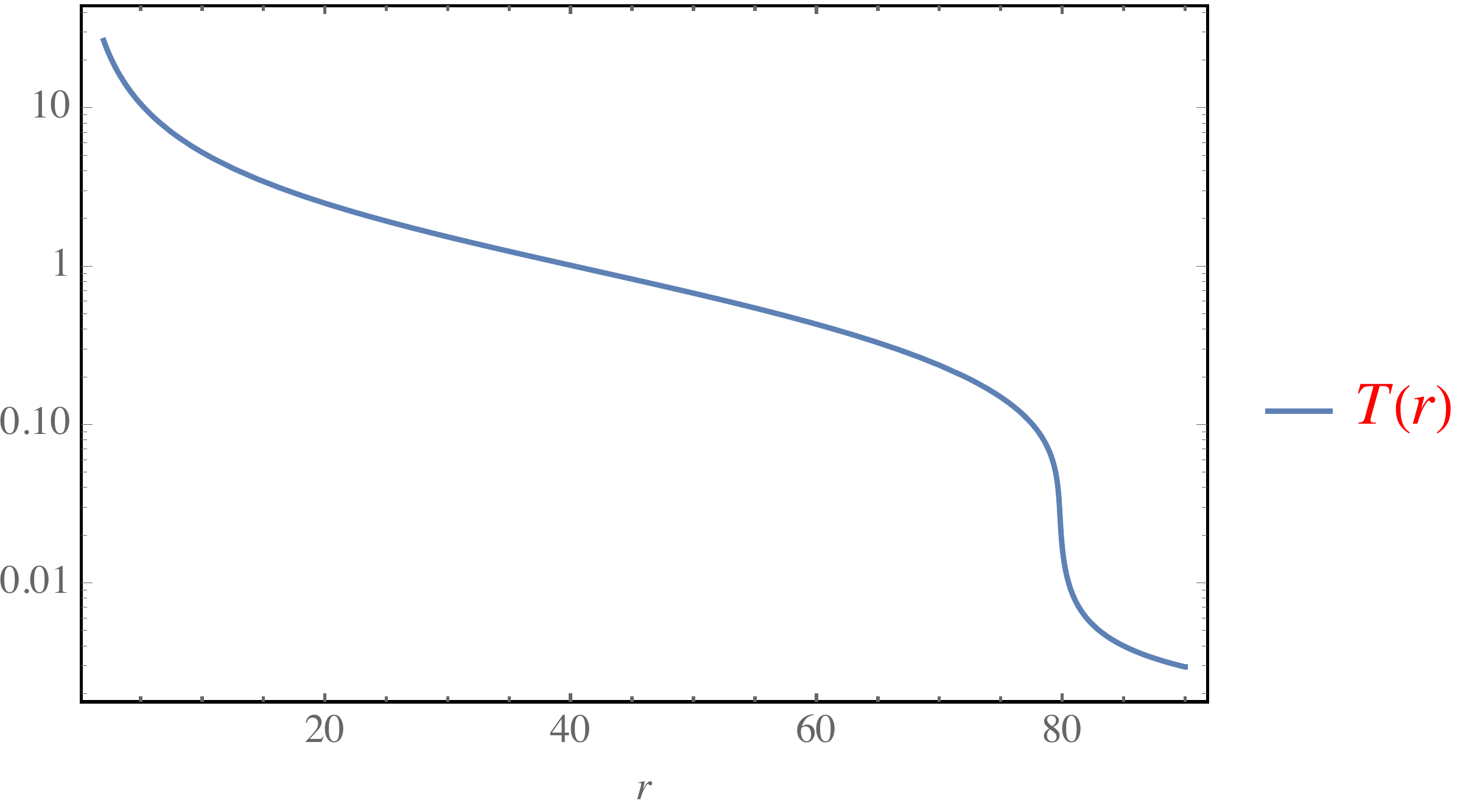}  
\caption{The local temperature $T(r)$ in Planck units and on a log scale.}
\label{fig6}
\end{figure}

\begin{figure}[!h]
\centering
   \includegraphics[width=9cm]{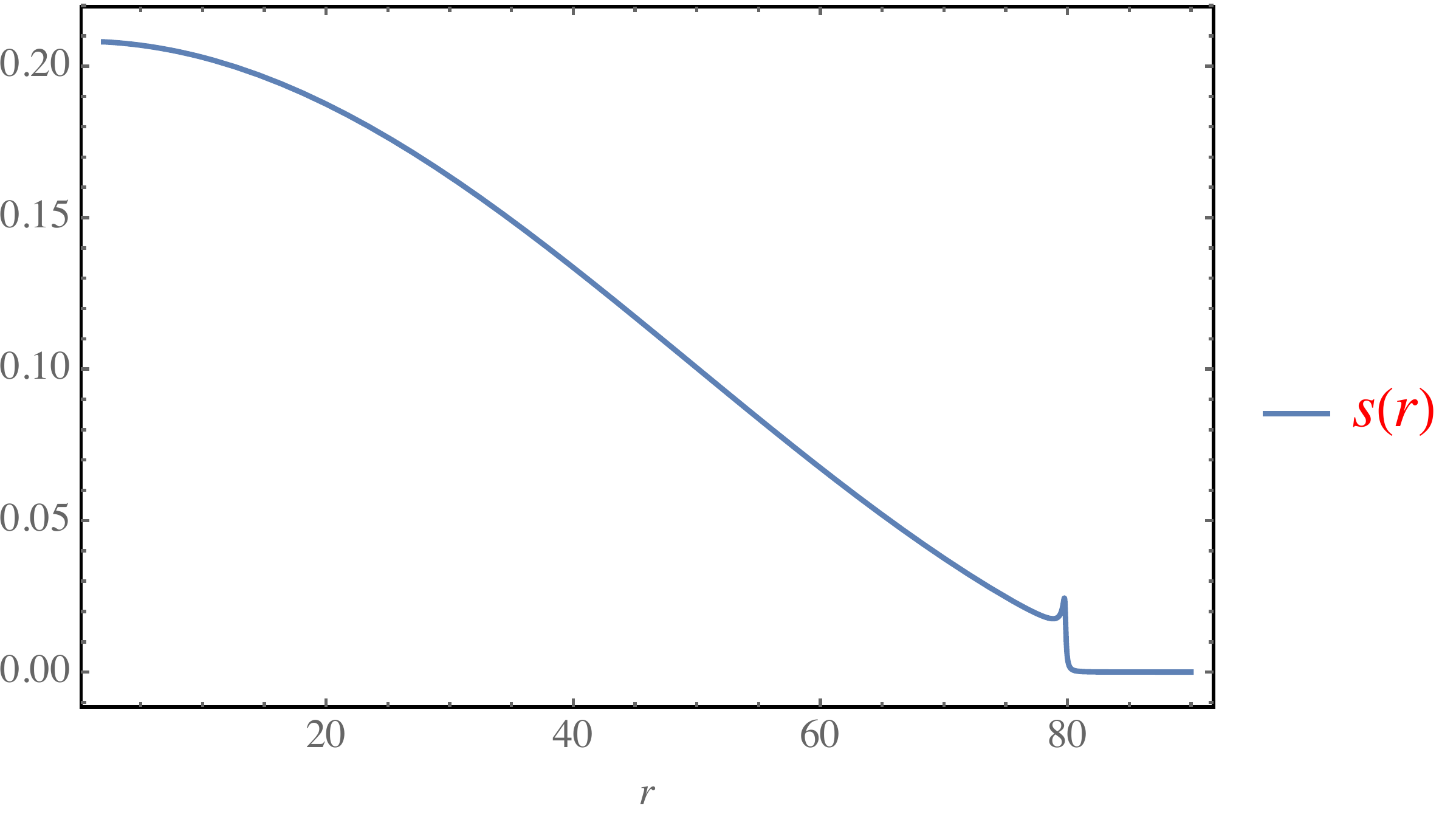}  
\caption{The entropy density $s(r)$, defined as $S=\int s(r)dr$. The area law for entropy comes not from having the entropy located on the surface, but from how $A(r)$ and $B(r)$ scale with $r_h$. The small spike at $r=r_h$ also grows with increasing $r_h$.}
\label{fig7}
\end{figure}

\begin{figure}[!h]
\centering
   \includegraphics[width=9cm]{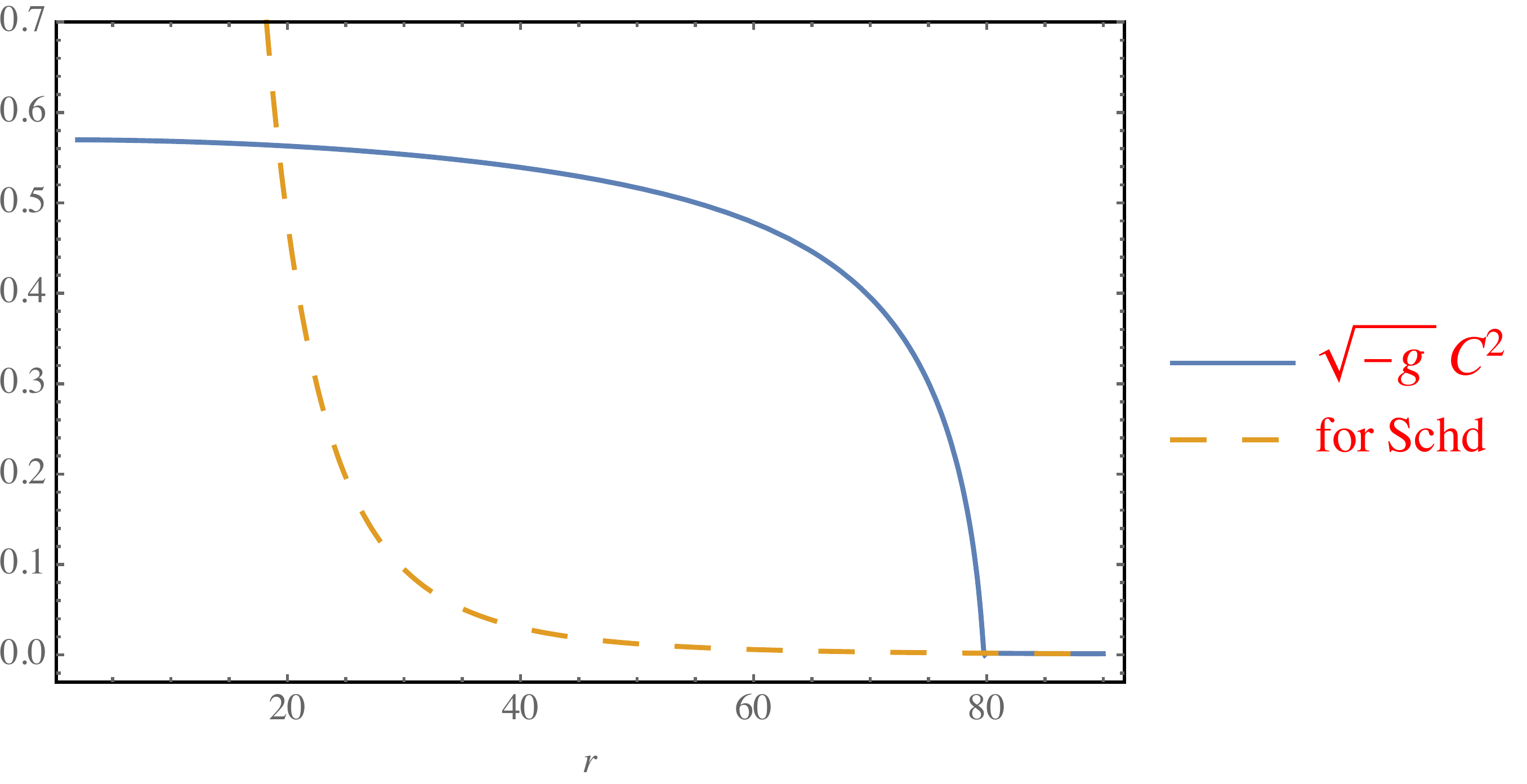}  
\caption{The action density. This shows that ``Weyl dynamics'' vs ``Einstein dynamics'' has softened the behaviour of this quantity.}
\label{fig8}
\end{figure}

Figures \ref{fig1} through \ref{fig8} illustrate the nature of these new 2-2-hole solutions. See the figure captions for details.


\begin{figure}[!h]
\centering
   \includegraphics[width=9cm]{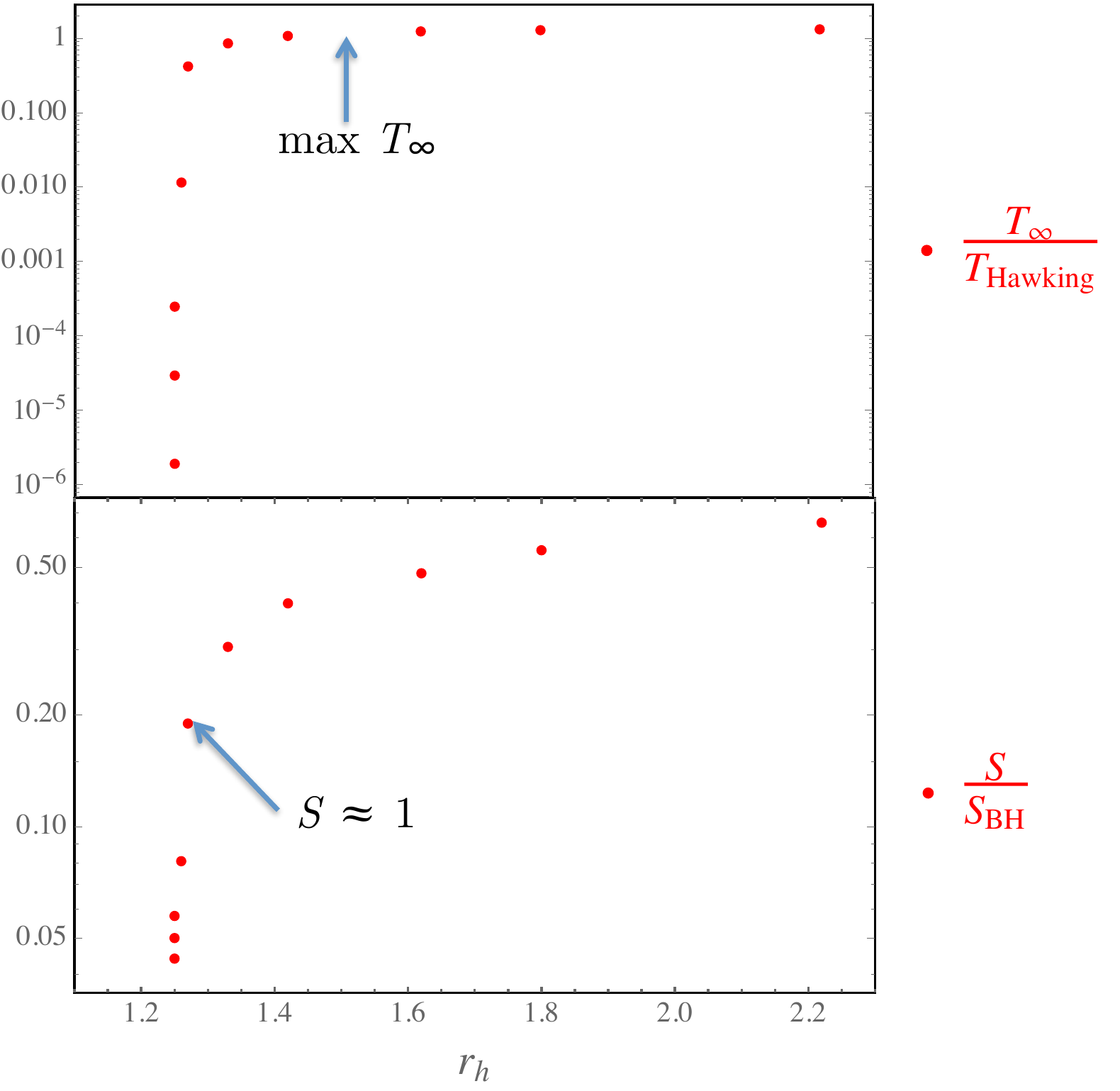}  
\caption{Approach to the smallest 2-2-holes. We set $N=1$ for display purposes.}
\label{fig9}
\end{figure}

We now have a look at small 2-2-holes. We find a minimum size of $r_h\approx1.25$. In Fig.~(\ref{fig9}) we show how $T_\infty$ and $S$ vary compared to black hole values as this minimum radius is approached. The maximum $T_\infty$ is reached at $r_h\approx1.5$. Thus for a $T_\infty$ less than this there are two 2-2-hole solutions, one large and one small. This is not unlike the situation in QCD, with hadrons and the quark matter solutions. We see that the minimum $r_h$ solutions are characterized by rapidly falling $T_\infty$ and more slowly falling $S$. But when $S$ approaches order one, there are no longer continuous values of $T_\infty$ and $S$. The minimum nonvanishing entropy corresponds to one particle trapped inside.

\begin{figure}[!h]
\centering
   \includegraphics[width=9cm]{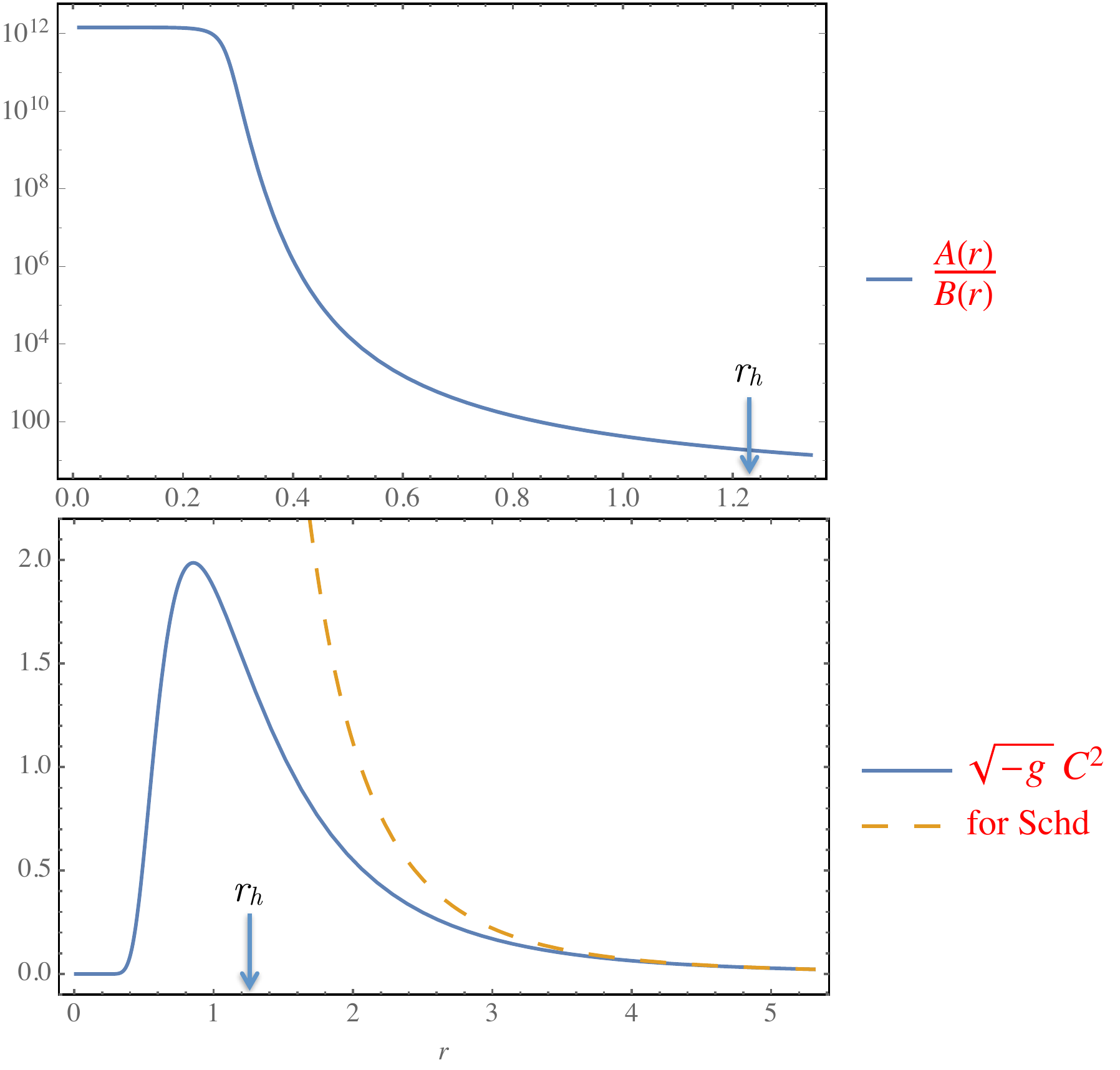}  
\caption{Illustrative quantities for the solution at the smallest $T_\infty$ point of the previous plots.}
\label{fig10}
\end{figure}

Fig.~(\ref{fig10}) shows some features of a small 2-2-hole, which differs markedly from a large 2-2-hole. $A(r)/B(r)$ near the origin keeps growing for decreasing $T_\infty$, implying near vanishing speed of light and thus increasing light crossing time.

\section{Gravitational waves}
\label{s5}

2-2-holes and their many novel properties are a reflection of the enlarged solution space of quadratic gravity. General relativity has not given us experience with such objects. But so far to the external observer, a 2-2-hole may appear quite indistinguishable from a black hole. In this section we show that this is not the case. Let us return to the large 2-2-holes and rewrite the previous wave equation using the tortoise coordinate $r_*$ (defined below) and $\psi_l(r,t)=e^{-i\omega t}\Psi_l(r)/r$ to give
\begin{align}
(\partial^2_{r_*}+\omega^2-V_l(r))\Psi_l=0,\quad\Psi_l(0)=0.
\end{align}
The behaviour of $\psi_l(r,t)$ near the origin as noted above translates into the Dirichlet boundary condition on $\Psi$. The information of the spacetime is encoded in the potential $V_l(r)$.

\begin{figure}[!h]
\centering
   \includegraphics[width=9cm]{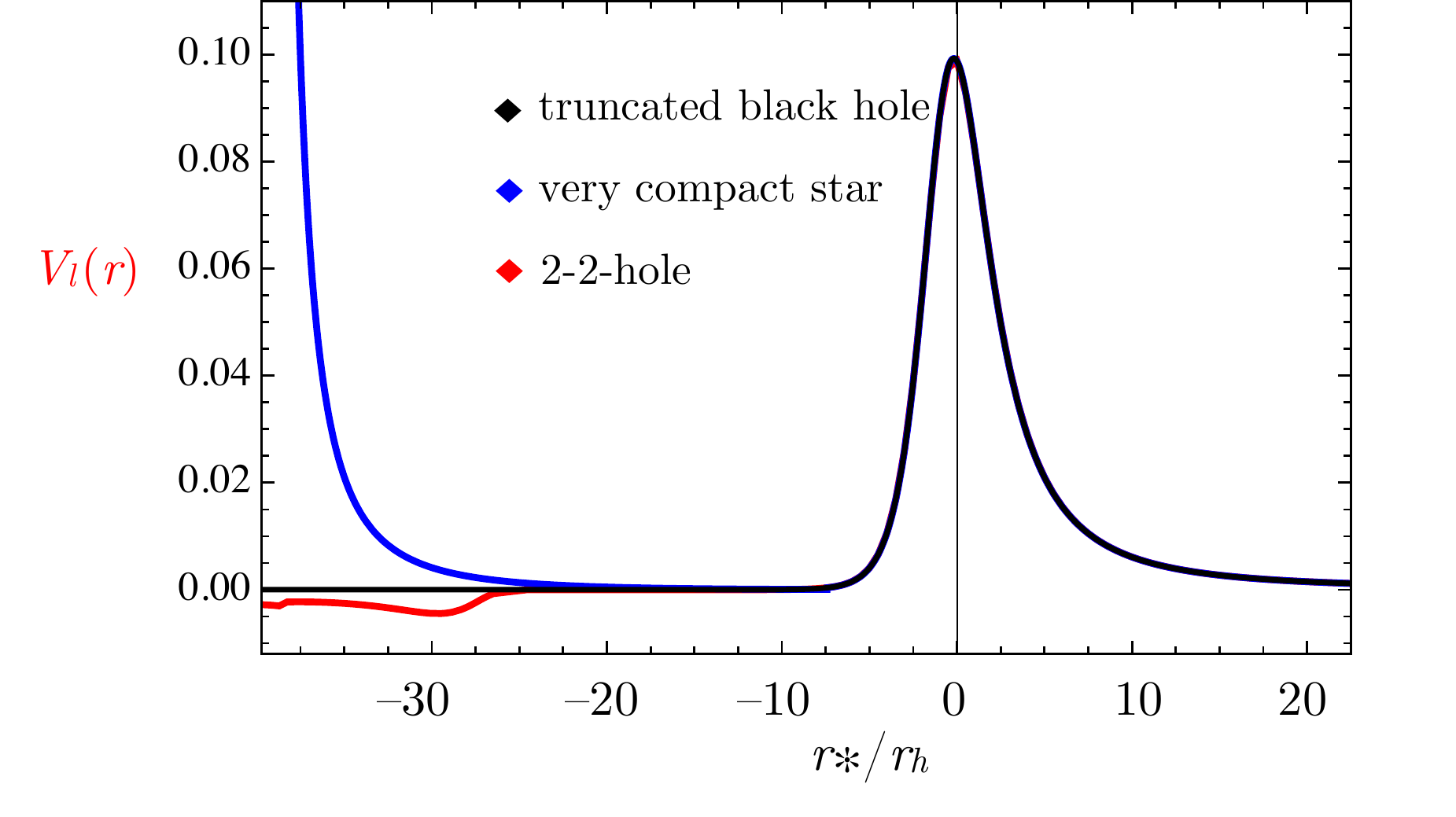}  
\caption{The potential as seen be scalar waves.}
\label{fig20}
\end{figure}

In Fig.~(\ref{fig20}) we compare this potential for three different objects.
All the three objects have the same potential barrier situated at $1.5r_h$. The very compact star displays the standard centrifugal barrier at the origin, while the 2-2-hole does not. The truncated black hole is an artificial but well studied construction (starting with the brick wall model) where the black hole geometry is modified by the placement of a wall just slightly outside the horizon. We thus see that a truncated black hole with a Dirichlet boundary condition is not a bad approximation to a 2-2-hole. We will find this to be a useful approximation to a spinning 2-2-hole, for which we don't yet have the explicit solution. The region to the left of the potential barrier can be considered a cavity, and the figure shows that a 2-2-hole potential deviates from the featureless truncated black hole in a region close to the the wall, that is the 2-2-hole origin. This region with a nontrivial potential close to the origin corresponds to the interior of the 2-2-hole. 

We now turn to the fact that a realistic ``cavity size'' is larger than shown in the figure and is closer to $\Delta r_*/r_h \approx 155$. This size determines a certain time delay of physical interest. Consider a spherical gravitational disturbance that starts at the potential barrier, travels inwards, encounters the reflecting boundary condition at the origin, and then travels back out to the potential barrier. The time it takes is
\begin{align}
\Delta t=2\Delta r_*=2\int_0^{\frac{3}{2}r_h}\sqrt{\frac{A(r)}{B(r)}}dr\approx 2r_h\textcolor{red}{\eta} \log(\frac{r_h}{\ell_{\rm Pl}}).
\label{e3}
\end{align}
Included here is the definition of the tortoise coordinate, which is a rescaled radial coordinate such that the radial speed of light reverts back to unity. 

The log enhancement of the time delay comes from the sharp peak in $A(r)/B(r)$ located just outside $r_h$. The time delay for a truncated black hole with a wall located at $r_{\rm peak}$ is similar, since the additional interior contribution from the 2-2-hole is relatively small. The smaller the distance $r_{\rm peak}-r_h$ is, the larger the peak height and the larger the $\eta$ factor. How the peak grows with $r_h$ was illustrated in Fig.~(\ref{fig4}), but the peak height for a 2-2-hole relevant to LIGO is not readily accessible. A rough extrapolation from much smaller spinless 2-2-holes in our previous work showed that $\eta\sim2$ was reasonable. The value $\eta\approx2$ (1) corresponds to $r_{\rm peak}-r_h$ being a proper (coordinate) Planck length.

We can now understand the basic behaviour of gravitational perturbations of a 2-2-hole. Consider low frequency gravity waves that have low transmission through the potential barrier. In this case we have a cavity with one end that slightly leaks. Then the ``quasi-normal-modes'' are the standing wave modes of the cavity that slowly decay due to the leak. An interesting observation by Cardoso and Pani (2017) is that a gravitational disturbance that bounces back and forth within the cavity gives rise to echoes on the outside. More generally, the frequency content of whatever leaks from the cavity (whether or not it gives well defined echoes) is dominated by the spectrum of QNMs.

The black hole mergers being observed by LIGO have rapidly spinning final states. Since spinning 2-2-hole solutions are not available, we model a spinning 2-2-hole by a Kerr black hole that is truncated just outside its horizon. When compared to the spinless truncated black hole, the QNM frequency spectrum is changed quite substantially. The phenomena of ``ergoregion instability'' and ``superradiance'' are also introduced, but these effects can be controlled by a small amount of dissipation of the gravity wave inside the 2-2-hole.

If LIGO is actually observing 2-2-hole merger events that each result in new 2-2-hole, then we want to know the signals produced by the final 2-2-hole as it relaxes to its ground state. The initial disturbance is reprocessed by the physics of the cavity, and this is most clearly described in frequency space by a transfer function. This arises in a Greens function method and the transfer function multiplies the frequency content of the initial perturbation to produce the frequency content observed at infinity.

\begin{figure}[!h]
\centering
   \includegraphics[width=9cm]{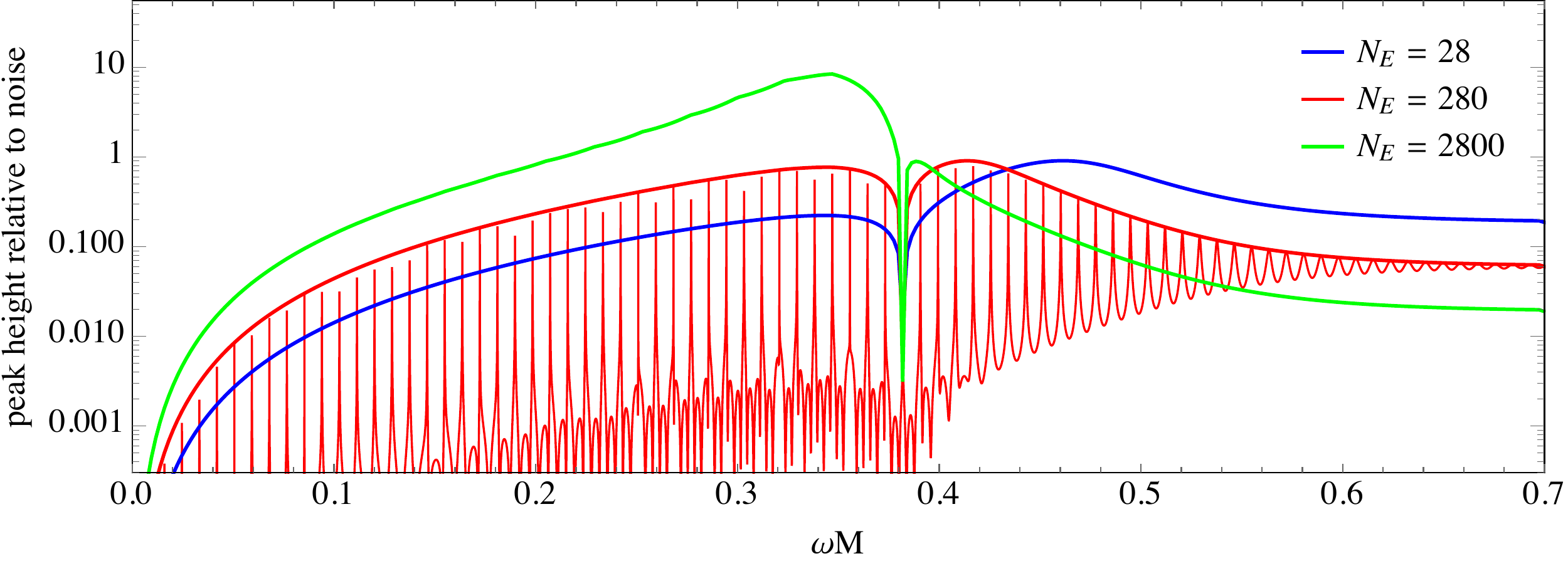}  
\caption{The absolute value of the transfer function for a spin of $2/3$. The location of an obvious special frequency is determined by the spin.}
\label{fig21}
\end{figure}

The spiky curve in Fig.~(\ref{fig21}) gives an example of the absolute value of the transfer function for particular range of data. The longer the time range (the more echoes $N_E$) the more resolution there is and the higher the very narrow spikes can appear. The different smooth envelope curves are also normalized by the noise that accumulates over longer times.
For much of the frequency range the transfer function is completely dominated by the spikes of the sharp QNM resonances. The frequency spacing of the resonances is the inverse of the time delay, and thus the large number of resonances is related to log enhancement of the time delay as discussed above.

A large number of equally spaced sharp resonances is a tempting target to search for, even for theorists! With LIGO data publicly available, we developed a search strategy as follows. Take the LIGO waveform for a range of time after the merger event, fourier transform, and then take the absolute value. Search for $\sim50$ evenly spaced spikes by multiplying the data with a comb-like function. This defines an amplitude as a function of the spacing and shift, and we then look for a correlated increase in the two amplitudes from the two detectors.

Note that an absolute value is being taken since this is what displays the simple resonance pattern. There is much more information in the full echo waveform, and its fourier transform, but this information is superfluous to the signal of interest. In contrast the LIGO collaboration prefers to use matched filter techniques on the full waveform. They are then forced to employ a large enough space of templates to model the full information content, which includes for example the relative phases between $\sim50$ spikes. But we have argued that the main feature of echoes lies in a much smaller space, and it advantageous to directly search in this smaller space.

For these reasons and others, LIGO has thus far not seriously searched for echoes. Thus we are in the suboptimal situation where outsiders are looking for echoes in the public data, even though only LIGO truly understands its own data. With these caveats we simply present what we find. And what we find is that there is structure in the data that is very consistent with an echo interpretation. There are a number of things that we look for in an echo signal. The location of the dominant frequency spikes should be consistent with the transfer function as in Fig.~(\ref{fig21}), which is controlled by mass and spin. The signal should have persistence over quite different time durations of data in a single event. In fact the strongest signals typically come from time series containing on the order of 150-200 echoes, which correspond to quite different time durations for different events. We also typically find secondary signals when the comb spacing is doubled or halved, a property which occurs less frequently for faked signals. And finally we check that the absolute location of the resonance pattern (the shift of the comb in frequency space) is a stable quantity.

Of most immediate interest is the time delay $\Delta t$ of each event, which is expected to be related to the mass of the final object from (\ref{e3}) ($r_h=2GM$). This relation should be generalized to the case of spin, and for a truncated Kerr black hole this relation is known,
\begin{align}
\Delta t=\eta r_h \log(\frac{\textcolor{red}{r_h}}{\ell_{\rm Pl}}) (1+1/\sqrt{1-\textcolor{red}{\chi}^2})(1+\textcolor{red}{z}).
\label{e5}\end{align}
We have commented that the interior of a spinless 2-2-hole gives a minor contribution to $\Delta t$, and here we assume the same for a spinning 2-2-hole. The three quantities on the RHS of (\ref{e5}) are measured by LIGO for the final state of each event ($\chi$ is the spin and $z$ is the redshift). When there is an echo signal, $\Delta t$ can be determined with relatively negligible error due to the large number of evenly spaced spikes. With echo signals from multiple events, then each can be considered to be a measurement of $\eta$. The consistency of the measurements of $\eta$ is then also important to establish the consistency of the echo interpretation.

\begin{figure}[!h]
\centering
   \includegraphics[width=9cm]{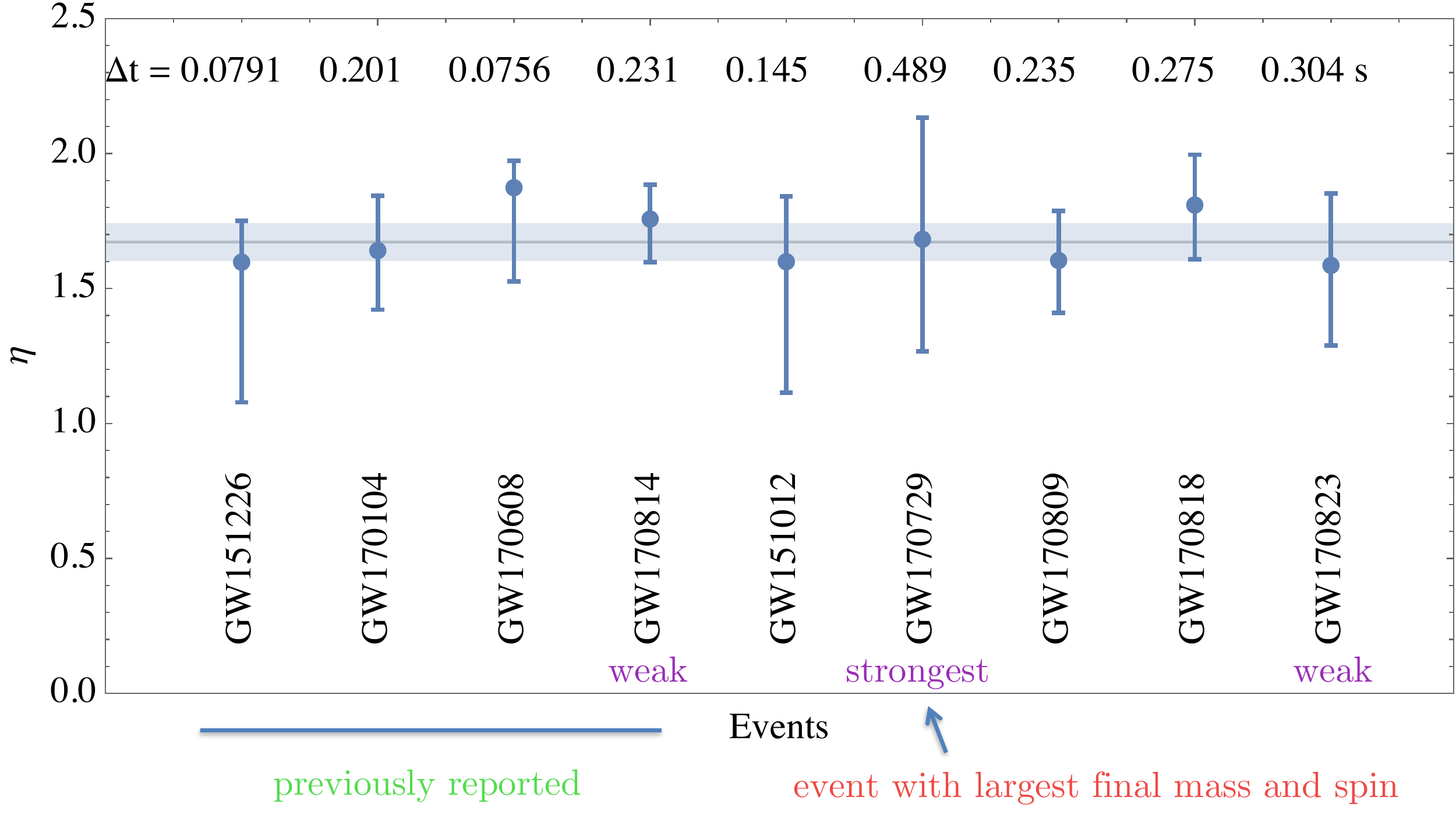}  
\caption{Nine black hole merger events, each with an echo signal of varying strengths. Time delay $\Delta t$ and the implied value of $\eta$ is shown for each.}
\label{fig22}
\end{figure}

Our results are shown in Fig.~(\ref{fig22}). Our previous work reported on four merger events, and here we include five more events for which public data appeared more recently. The obtained $\eta$'s are very consistent. The error bar of each is due to the errors in the LIGO measurements of $r_h$, $\chi$ and $z$. The strength of the echo signal in each event cannot be seen from this plot, although a few of the strengths are indicated.
As we have described, the measurement $\eta\approx1.7$ is a measurement of $r_{\rm peak}-r_h$, and the result can be expressed in two ways. This distance is $\approx10^{-28}\times\ell_{\rm Pl}$ or $\approx10^{12}\times(\mbox{proper Planck length})$.

\section{Conclusions}
Ghosts and naked singularities are problems that are often encountered in theories of gravity. Intuition from classical theories suggest that these problems are fatal, since the first means that time evolution is unstable, while the second means that time evolution is ambiguous. Intuition about these problems in quantum gravity is much less developed. Quadratic gravity offers a quantum theory that has close similarities to an asymptotically free gauge theory where the Planck scale can arise as a dynamically generated mass scale. We find that the ghost transforms into a slight acausal behavior of the full graviton propagator.

At the same time macroscopically large solutions can be well studied by the classical theory. A horizonless spacetime with a naked timelike singularity emerges, but the latter is of a type that does not lead to any ambiguous evolution of fields. We have reported here on new 2-2-hole solutions that are sourced by a relativistic gas. We find an area law for the entropy that can be larger than the same size black hole. And in contrast to black holes, there is a clear understanding of the entropy and its ties to the geometry.

\appendix*
\section{Thoughts on the cosmological constant}

Consider a theory with classical scale invariance and consisting of asymptotically free gauge interactions only. That is, no elementary scalars or dimensionful parameters are present. In such a theory the CC is thought to be generated through dimensional transmutation and be of order $\Lambda^4$, where $\Lambda$ is the highest such scale. But how can we be sure of this? Even the meaning of $\Lambda$ is uncertain. With the standard definition,
\begin{align}
\log \frac{\Lambda}{\mu}=\int_{g(\mu)}^\infty\frac{dx}{\beta(x)},
\end{align}
it is implicitly assumed that there is an IR Landau pole. If nonperturbative effects cause the effective coupling to saturate at some finite value, then the implied $\Lambda$ from this effective coupling vanishes. The theory certainly generates a mass gap, but we claim that it is not certain that it generates a CC.

The vacuum energy in a QCD-like theory with no current masses (massless QCD), is related to the existence of a gluon condensate by the trace anomaly. The gluon condensate in turn is related to the existence of certain power law corrections in the UV, via the operator product expansion. In the lattice formulation of the theory it may be easier to use these power law corrections to infer the gluon condensate rather than trying to extract the gluon condensate directly. This was demonstrated by D'Elia, Giacomo, Meggiolaro (1997), but unfortunately this approach has not been pursued. Thus we are still left wondering, could power law corrections in massless QCD be softer in the UV than naively expected? Does a gluon condensate actually exist in massless QCD? Real QCD doesn't answer the question since the strange quark mass fakes a gluon condensate.

One picture of nonperturbative effects in gauge theories originates in the measure of the path integral, a phenomenon known as ``Gribov copies''. These effects are very nonlocal and the resulting UV corrections are exponentially soft. Thus these effects are not a source of power law corrections or condensates. But they can still be the source of a mass gap, and in fact there is a whole picture of the mass gap and confinement based on Gribov copies. (And in turn this mass gap causes the coupling to saturate.) We are raising this picture to further suggest that generating a mass gap and generating a CC is not necessarily the same thing. If this possible loop-hole is realized in nature then all masses have a dynamical origin and they all, eventually, die away in the UV, allowing UV power law corrections to be softer than what a CC would imply. This picture also points us towards the idea that the Planck mass is just another dynamically generated mass.

\begin{acknowledgments}

I would like to thank the organizers of the CERN workshop ``Scale Invariance in Particle Physics and Cosmology'' for a fruitful gathering. I thank Jing Ren for useful discussions. This research is supported in part by the Natural Sciences and Engineering Research
Council of Canada. 

\end{acknowledgments}	
%
\end{document}